\documentclass[aps,pra,reprint,groupedaddress,tightenlines,nofootinbib]{revtex4-2}
\usepackage{amssymb}
\usepackage{color,graphicx}
\usepackage{xcolor}
\usepackage{lmodern}
\usepackage{amsmath}
\usepackage{amsbsy}
\usepackage{float}
\makeatletter
\let\newfloat\newfloat@ltx
\makeatother
\usepackage{bbm}
\usepackage[most]{tcolorbox}
\usepackage{bm}
\usepackage{epsfig}
\usepackage{braket}
\usepackage{graphics}
\usepackage{empheq}
\usepackage{lipsum}
\usepackage{mathrsfs}
\usepackage{physics}
\usepackage{mathtools}
\usepackage{algorithm}
\usepackage{algorithmic}
\floatstyle{plain}
\restylefloat{algorithm}
\usepackage[colorlinks=true,citecolor=blue!65!black,linkcolor=blue!65!black,urlcolor=blue!65!black]{hyperref}
\renewcommand{\thealgorithm}{\arabic{algorithm}}
\usepackage{booktabs}

\newcommand{\Nshot}{N_{\mathrm{shot}}}

\newcommand{\AIC}{\mathrm{AIC}}
\newcommand{\BIC}{\mathrm{BIC}}

\begin{document}
\title{Tomographic reconstruction of non-Markovian Gaussian noise in continuous-variable photonic systems}
\author{Hamid Reza Naeij}
\email[]{hnaeij@mail.uni-paderborn.de}
\affiliation{Institute for Photonic Quantum Systems (PhoQS) and Department of Computer Science, Paderborn University, Germany}
\begin{abstract}
Memory effects in continuous-variable (CV) photonic systems can make standard Markovian Gaussian noise models inaccurate, even when the average noise strength is correctly calibrated.  We propose a theoretical and simulation-based tomography framework for reconstructing such memory effects directly from time-resolved homodyne data. 
The method estimates an effective time-local Gaussian generator, represented by drift and diffusion matrices, using a regularized maximum-likelihood procedure constrained by complete positivity (CP) of the finite-time Gaussian channel. This description covers any Gaussian-preserving
noise dynamics for which a well-defined time-local generator
exists, rather than being restricted to particular noise
mechanisms.
Non-Markovianity is then diagnosed at the generator level through temporary violations of local CP-divisibility, rather than from time dependence alone. This distinction is important because ordinary nonstationary noise can remain
CP-divisible, whereas genuine memory can produce a non-CP-divisible reduced generator while the accumulated channel remains physical. 
Numerical benchmarks covering memory-dependent loss, nonstationary thermal excitation, and pseudomode-mediated memory show that the reconstruction distinguishes CP-divisible time-dependent noise from genuine non-CP-divisible dynamics and captures the corresponding signatures in the Gaussian covariance evolution. The framework therefore provides a practical route for diagnosing non-Markovian Gaussian noise in CV photonic quantum systems.
\end{abstract}
\maketitle
\noindent{\textbf{Keywords}: Continuous-variable quantum systems, Non-Markovian dynamics, Gaussian quantum channels,
Quantum tomography, Homodyne detection,
CP-divisibility.}

\section{Introduction}
CV photonic systems are a leading platform for quantum information, quantum communication, quantum computing, sensing, and simulation \cite{Serafini2017,Zhang2019,Usenko2026,Rudolph2025Energy,Lloyd1999}. In these systems, information is encoded in the quadratures of optical modes,
which can be prepared as coherent or squeezed states, transformed by Gaussian operations, and measured using homodyne detection \cite{Weedbrook2012,Braunstein2005,Adesso2014,Ferraro2005}. In many CV photonic systems, the relevant dynamics is well described in the Gaussian regime, where the state is characterized by its displacement vector and covariance matrix. Noise analysis in this setting is commonly formulated in terms of Gaussian
channels, including photon loss, thermal noise and additive Gaussian
noise \cite{Holevo2007,Gardiner2004,Hastrup2021}. These models are powerful because Gaussian noise can be represented compactly via drift and diffusion matrices that govern the evolution of the first and second moments \cite{Liu2024,Mirkhalaf2024,Ernesto2026,Massignan2015,Ferraro2005,Ma2018}.

Most standard noise models for CV photonic systems assume
a memoryless environment and describe CP-divisible evolution
\cite{Breuer2007,Schlosshauer2007}. A physical density
operator is positive semidefinite and has unit trace,
ensuring non-negative measurement probabilities that sum
to one. A trace-preserving quantum dynamical map $\Lambda$
is completely positive (CP) if it maps every physical
density operator to a physical density operator, including
when applied to one part of an entangled composite system
while leaving the other part unchanged. For example,
attenuating both quadratures of the vacuum without adding
the required noise would violate the Heisenberg uncertainty
relation. Such a prediction indicates an inconsistent
model, not a realizable deterministic process. The evolution
is CP-divisible if every intermediate map is also CP and trace-preserving, namely
\cite{Gorini1976,Chruscinski2011}
\begin{equation}
\begin{aligned}
&\Lambda(t,0)=\Lambda(t,s)\Lambda(s,0)\\
\qquad
&\Lambda(t,s)\ \text{is CPTP for all } t\ge s\ge 0
\end{aligned}
\end{equation}
where CPTP denotes Completely Positive and Trace Preserving dynamics.
\begin{figure*}
\centering
\includegraphics[scale=0.99]{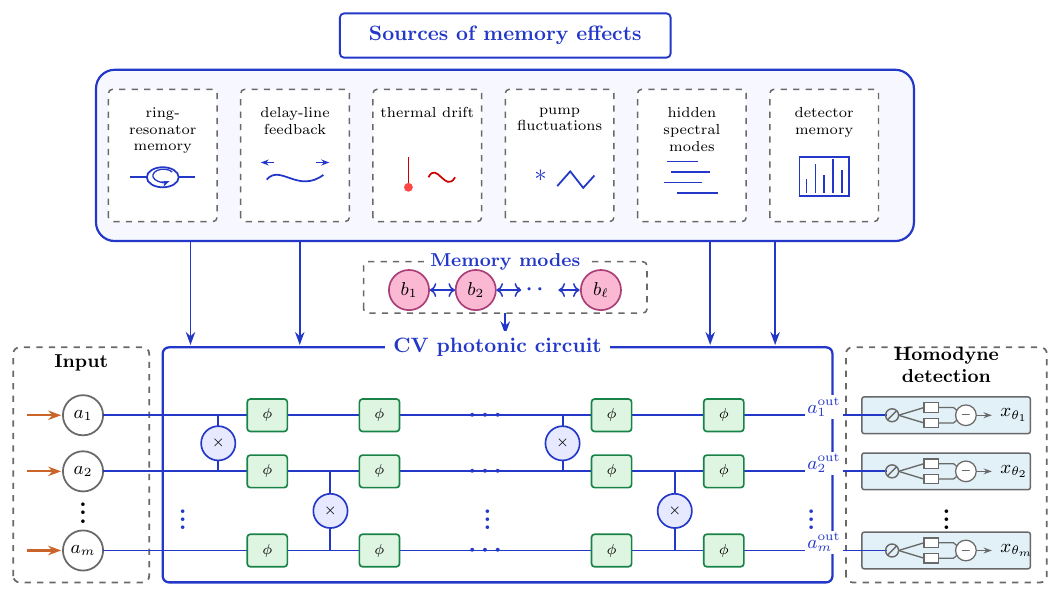}
\caption {Schematic of a CV photonic circuit subject to possible memory-bearing noise sources. Resonator modes, delay-line feedback, temporally correlated thermal drift, pump fluctuations and hidden spectral modes can generate time-correlated system noise, whereas detector memory can introduce correlations in the measured homodyne record. Auxiliary pseudomodes represent environmental memory and possible information backflow. These mechanisms produce non-Markovian reduced dynamics only when the reconstructed time-local generator violates CP-divisibility while the finite-time Gaussian channel remains CP. Homodyne detection provides the quadrature data used for generator-level tomography.}
\label{fig:cv_memory_schematic}
\end{figure*}

Realistic hardware, however, can contain temporally correlated Gaussian noise sources, including resonator memory, delay-line feedback, thermal drift, pump-amplitude and pump-phase fluctuations, hidden spectral modes, correlated dissipation and detector-induced correlations \cite{Gardiner2004,Hastrup2021}. 
Non-Markovian open quantum system theory provides a natural language for describing such memory effects, especially through time-local Lindblad-like master equations with time-dependent coefficients.
If the instantaneous generator violates the condition for
CP-divisibility while the total map remains CP, the dynamics
is non-Markovian in the divisibility sense
\cite{Breuer2007,Breuer2009,Gorini1976,Rivas2014,Schlosshauer2007,Chruscinski2013,Chruscinski2011,Chruscinski2012}.
Physically, the environment can retain information about
the system's earlier evolution through system-environment
correlations. Consequently, an intermediate map need not
represent a physical operation on an arbitrary system state
independently of these correlations. Its failure to be CP
does not make the actual evolving state unphysical:
the total map from the initial time remains CP. Throughout this work, we focus on non-Markovianity characterized by violations of CP-divisibility, which provides a
generator-level diagnostic complementary to broader
descriptions based on multi-time correlations \cite{Milz2019}.
Accordingly, CP-divisibility alone does not exclude
such correlations.

In CV Gaussian systems, memory effects have been investigated through several complementary diagnostics. Fidelity-based distinguishability measures adapt the information-backflow picture to Gaussian dynamical maps and have been applied to damping and quantum Brownian-motion channels. Divisibility-based approaches provide criteria for non-Markovianity directly at the level of Gaussian channels by testing the CP of intermediate maps, while related Gaussian-process hierarchies distinguish Markovian, weakly non-Markovian, and strongly non-Markovian regimes. Beyond channel-level diagnostics, structured reservoirs and fluctuating environments can lead to non-monotonic behavior of CV resources such as two-mode entanglement, logarithmic negativity, quantum discord, intensity correlations, Gaussian interferometric power and revivals of quantum correlations \cite{Vasile2010,Vasile2011,Vasile2009,Torre2015,Souza2015,Trapani2016,LiuzzoScorpo2017}. These results motivate a generator-level tomographic reconstruction strategy in which memory is identified not only through state-dependent witnesses, but also through the reconstructed time-local drift and diffusion matrices.

Tomographic methods provide a well-established framework
for reconstructing unknown quantum states and processes
from measurement data \cite{DAriano2024,Merkel2013}.
In CV systems, homodyne tomography provides direct access
to quadrature statistics and enables the reconstruction
of Gaussian states, covariance matrices, and Gaussian
channels \cite{DAriano1995,DAriano1999,DAriano1994,Lvovsky2009}.
Maximum-likelihood and constrained estimation methods
provide statistical tools for quantum tomography
\cite{James2001,Banaszek1999,Fiurasek2001,Sacchi2001},
including state estimation with additive Gaussian
measurement noise \cite{Smolin2012}. Related applications
of likelihood-based inference include amplitude estimation
on noisy quantum computers \cite{Tanaka2021} and noise
estimation for quantum error correction \cite{Cao2026}.
For non-Markovian dynamics, process tomography can
characterize multi-time correlations \cite{White2022}.
Most closely related to our methodology, Varona
et al. \cite{Varona2025} developed Lindblad-like quantum
tomography to reconstruct time-local master equations,
including temporarily negative decay rates, through
constrained maximum-likelihood estimation for qubit
dynamics. Aguiar et al. \cite{Aguiar2025} introduced
regression-based quantum Liouvillian tomography and
applied it to superconducting-qubit data, while
Reddy \cite{Reddy2025} investigated structure-preserving
reconstruction of non-Markovian time-local generators
using basis expansions and neural networks.
Related work by Guan et al. \cite{Guan2026} examined
the identifiability of drift and diffusion matrices
from temporal distributions of classical linear stochastic
systems. Complementarily, Torre et al. \cite{Torre2015}
established CP-divisibility criteria for Gaussian
quantum channels, with further characterization of
Gaussian non-Markovian dynamics and their classification
\cite{LiuzzoScorpo2017,Torre2018}. Renault
et al. \cite{Renault2023} experimentally characterized
environmental memory and non-Markovian effects using
squeezing and entanglement correlations in a CV optical
platform. 

Although existing Gaussian-channel criteria provide
a rigorous characterization of CP-divisibility,
their integration with data-driven reconstruction
of time-local Gaussian generators remains an important
methodological problem. Conversely, likelihood-based
generator reconstruction has primarily been developed
for finite-dimensional quantum systems, without
explicitly addressing the drift-diffusion structure
and physicality constraints of Gaussian CV dynamics.
The central challenge is therefore to reconstruct
time-local Gaussian generators from time-resolved
homodyne statistics while enforcing CP
of the finite-time evolution and independently assessing
the CP-divisibility of the reconstructed dynamics.

In this work, we address this challenge by developing
a theoretical and simulation-based framework for
tomographic reconstruction of non-Markovian Gaussian
noise in CV photonic systems.
The unknown dynamics is represented by a time-local Gaussian generator with drift matrix, diffusion matrix, and displacement drift. 
These quantities determine the evolution of the displacement
vector and covariance matrix, which completely specify
each output Gaussian state and hence its homodyne
probability distributions. The dynamics considered here preserves Gaussian states
in both the Markovian and non-Markovian regimes.
We estimate the generator parameters using a regularized maximum-likelihood procedure constrained by CP of the total Gaussian channel. Non-Markovianity is then diagnosed by testing whether the reconstructed infinitesimal dynamics violates CP-divisibility. Figure \ref{fig:cv_memory_schematic} summarizes the physical picture considered
in this work: a CV photonic circuit is treated as the accessible system, while memory-bearing environmental degrees of freedom can produce time-correlated loss, diffusion, and correlated noise that are reconstructed from homodyne data. 

The paper is organized as follows. Section \ref{sec:framework} introduces the theoretical framework for describing Gaussian Markovian and non-Markovian dynamics in CV photonic systems. Section \ref{sec:noise} connects this framework to physical noise sources and memory mechanisms. Section \ref{sec:tomogtaphy} presents the homodyne-based tomographic reconstruction method. Section \ref{sec:memory_detection} develops the diagnostic and model-selection procedures used to identify non-Markovian memory. Section \ref{sec:numerical_validation} provides a numerical validation of the proposed approach. Finally, Section \ref{sec:conclusion} summarizes the main results and outlines future directions.

\section{Theoretical Framework}
\label{sec:framework}
\subsection{Quadrature representation of \\CV photonic systems }
We consider an $m$-mode bosonic system with annihilation and creation operators \footnote{For notational simplicity, operator hats are omitted throughout; all mode, quadrature, and Hamiltonian symbols should be understood as operators unless explicitly stated otherwise.}
\begin{equation}
\bm{a}=(a_1,\ldots,a_m)^T,
\qquad
[a_j,a_k^\dagger]=\delta_{jk}
\end{equation}
Define quadrature operators
\begin{equation}
q_j = \frac{a_j+a_j^\dagger}{\sqrt{2}},
\qquad
p_j = \frac{a_j-a_j^\dagger}{i\sqrt{2}}
\label{eq:quadrature}
\end{equation}
and collect them into the quadrature vector
\begin{equation}
\bm{R}=(q_1,p_1,\ldots,q_m,p_m)^T
\label{eq:quadrature2}
\end{equation}
They obey
\begin{equation}
[R_j,R_k]=i\Omega_{jk}
\end{equation}
where the symplectic matrix is defined by
\begin{equation}
\Omega= \bigoplus_{j=1}^{m}
\begin{pmatrix}
0 & 1 \\
-1 & 0
\end{pmatrix}
\label{eq:symplectic}
\end{equation}

For Gaussian states, the first moment (displacement) and covariance matrix are defined respectively 
\begin{equation}
d_j(t)=\langle R_j\rangle_t
\label{eq:first_moment}
\end{equation}
\begin{equation}
V_{jk}(t)=\frac{1}{2}\langle\{R_j-d_j(t),R_k-d_k(t)\}\rangle_t
\label{eq:covariance}
\end{equation}
where the covariance matrix is a $2m \times 2m$ real and symmetric matrix which must satisfy the uncertainty principle \cite{Simon1994}
\begin{equation}
V(t)+\frac{i}{2}\Omega \geq 0
\end{equation}
The coherent Gaussian dynamics of such a photonic system is generated by a quadratic Hamiltonian 
\begin{equation}
\label{eq:Hamiltonian}
H(t)=\frac{1}{2}\bm{R}^T G(t)\bm{R}
\end{equation}
where $G(t)=G^T(t)\in \mathbb{R}^{2m\times 2m}$ is the real symmetric matrix generating beam splitters, phase shifters, squeezers, and mode mixing \cite{Weedbrook2012,Adesso2014}.
In addition to the quadratic Hamiltonian in Eq.\eqref{eq:Hamiltonian},
coherent driving is included through the linear term
$\mathbf{h}(t)^T\mathbf{R}$, where $\mathbf{h}(t)$ is
a real driving vector, generating the displacement
drift $\bm{u}(t)=\Omega\mathbf{h}(t)$ introduced below.

\subsection{Markovian Lindblad master equation}
The Markovian evolution of a photonic system coupled to a memoryless environment is described by the Lindblad master equation  \cite{Lindblad1976,Naeij2020,Naeij2025}
\begin{align}
\frac{d\rho(t)}{dt}&=
\mathcal{L}\rho(t)\nonumber=-i[H,\rho(t)]+
\sum_j \gamma_j \mathcal{D}[L_j]\rho(t)
\nonumber\\
&=-i[H,\rho(t)]+\sum_j \gamma_j\left(L_j \rho(t) L_j^\dagger-\frac{1}{2}\left\{L_j^\dagger L_j,\rho(t)\right\}\right)
\label{eq:lindblad}
\end{align}
Here $\rho$ is the reduced density matrix of the CV photonic system, $H$ is the system Hamiltonian, $L_j$ are Lindblad jump operators,
$\gamma_j\geq 0$ are relaxation rates or decay rates corresponding to each dissipative channel,
$\mathcal{D}[L_j]$ is the Lindblad dissipator associated with $L_j$, and $\mathcal{L}$ denotes the full Lindblad generator acting on $\rho(t)$. The Markovian model assumes that the environment has no memory and the instantaneous coefficients are time-independent or externally prescribed
non-negative functions.

For standard CV photonic hardware, the most common single-mode linear Lindblad operators describe photon loss and incoherent gain \cite{Gardiner2004,Hastrup2021,Breuer2007}
\begin{equation}
L^{\rm loss}_j=a_j,
\qquad
L^{\rm gain}_j=a_j^\dagger 
\label{eq:single_mode_linear_lindblad}
\end{equation}
More generally, Gaussian-preserving correlated dissipation and gain can be described by Lindblad operators that are linear combinations
of creation and annihilation operators
\begin{equation}
L_\mu=\sum_{j=1}^{m}\left(c_{\mu j}a_j+d_{\mu j}a_j^\dagger\right)
\qquad \mu=1,\ldots,N_L
\label{eq:linear_gaussian_lindblad}
\end{equation}
where the coefficients $c_{\mu j}$ and $d_{\mu j}$ determine which photonic modes $j$ are coupled to the dissipative environmental channel $\mu$. The special case
$d_{\mu j}=0$ corresponds to correlated photon loss, while nonzero $d_{\mu j}$ also allows gain- or squeezing-type Gaussian noise. Equivalently, correlated photon loss may be written in Kossakowski
form as introduced below in Eq.\eqref{eq:corr_loss_revised}.
Other important photonic noise mechanisms are generated by nonlinear Lindblad operators, for example
\begin{equation}
L^{\rm deph}_j=a_j^\dagger a_j,
\qquad
L^{\rm abs}_j=a_j^2 
\label{eq:nonlinear_lindblad_examples}
\end{equation}
which describe phase diffusion and two-photon absorption, respectively. They are useful physical noise models, but they generally go beyond the
strictly Gaussian-preserving drift-diffusion formulation used below, unless treated perturbatively or approximated within an effective
Gaussian model.

\subsection{Lindblad-like non-Markovian master equation}
Some non-Markovian dynamics can be described by a time-local master equation when a well-defined time-local generator exists. Such equations can be obtained, for example, from the time-convolutionless projection-operator formalism, and may be regarded as time-dependent generalizations of the usual Markovian Lindblad equation \cite{Naeij2025}. In this description, the Hamiltonian, dissipative operators, and decay rates can depend explicitly on time. If the instantaneous dissipative rates remain non-negative at all times, the generator has a Lindblad form at each instant and the corresponding dynamics is CP-divisible \cite{Breuer2007,Breuer2009,Rivas2014,Schlosshauer2007}. Here, CP is stronger than positivity of the reduced system alone: it guarantees that the evolution remains physically valid even when the system is initially correlated with an arbitrary ancilla \cite{Breuer2007}.
In contrast, if some instantaneous rates become temporarily negative while the overall dynamical map remains CP, the evolution is no longer Markovian in the CP-divisibility sense. This motivates a Lindblad-like time-local master equation
in which non-Markovian dynamics can be described by
temporarily negative canonical decay rates, while the
overall dynamical map remains CP
\cite{Breuer2009,Rivas2014,Hall2014}.
The canonical rates, namely the eigenvalues of the
dissipative coefficient matrix, are obtained by
diagonalizing this matrix in an orthonormal basis
of traceless operators; negative coefficients in an
arbitrary Lindblad representation do not by themselves
establish CP-indivisibility. For the Gaussian dynamics
considered here, we use the matrix criterion
$M_{\mathrm{CP}}(t)$ introduced below.

Here, we model the reduced dynamics of the photonic modes using an effective time-local master equation \cite{Rivas2014}
\begin{equation}
\frac{d\rho(t)}{dt}=\mathcal{K}(t)\rho(t)
\label{eq:time-local1}
\end{equation}
whose generator $\mathcal{K}(t)$ contains time-dependent Hamiltonian and dissipative contributions
\begin{equation}
\begin{aligned}
\mathcal{K}(t)\rho(t)&= -i[H(t),\rho(t)]\\
&+\sum_{\mu,\nu}C_{\mu\nu}(t)\left(F_\mu \rho(t) F_\nu^\dagger-\frac{1}{2}\{F_\nu^\dagger F_\mu,\rho(t)\}\right)
\label{eq:time-local2}
\end{aligned}
\end{equation}
where $\{F_\mu\}$ is a fixed, linearly independent family of noise operators and $C_{\mu\nu}$ are the entries of the time-dependent Kossakowski-like matrix $C(t)=C^\dagger(t)$ that specifies how strongly each pair of noise operators $(F_{\mu}$,$F_{\nu})$ contributes to the dissipative dynamics \cite{Kossakowski1,Kossakowski2}.
If
\begin{equation}
C(t)\geq 0 \quad \forall t
\end{equation}
then the dynamics is CP-divisible and Markovian in the divisibility sense. If $C(t)$ has negative eigenvalues at some times, the generator is not of standard Lindblad form at those times. Nevertheless, the total map
\begin{equation}
\rho(t)=\Lambda(t,0)\rho(0)
\end{equation}
may remain CP. Thus, non-Markovianity is indicated by
\begin{equation}
\exists t,\quad \lambda_{\min}[C(t)]<0
\label{eq:non-Markovianity}
\end{equation}
where $\lambda_{\min}[C(t)]$ denotes the smallest eigenvalue of the matrix $C(t)$. A negative value of $\lambda_{\min}[C(t)]$ indicates that $C(t)$ is not positive semidefinite, which signals a breakdown of CP-divisibility and is used as a witness of non-Markovianity of the quantum evolution.
Therefore, we call Eq.\eqref{eq:time-local1} a \emph{Lindblad-like non-Markovian master equation} \cite{Rivas2014,Breuer2009}.

\subsection{Gaussian-preserving Lindblad-like dynamics}
We focus on Gaussian-preserving dynamics, a regime relevant
to experiments such as squeezed-state-enhanced interferometry,
where squeezed light undergoes linear optical transformations
and Gaussian noise before homodyne detection. In this regime,
the state is fully characterized by its displacement vector
and covariance matrix, allowing us to describe the evolution
in phase space rather than through the full density operator.

In the phase-space representation, the quantum state is described by its Wigner function $W(\bm x,t)$, where the classical phase-space coordinate is
\begin{equation}
\bm{x}=(q_1,p_1,\ldots,q_m,p_m)^T\in\mathbb{R}^{2m}
\end{equation}
where $q_j,p_j$ are real phase-space coordinates
corresponding to the quadrature operators.
For the Gaussian-preserving models considered below, we assume that the Hamiltonian is at most quadratic in the quadrature operators and
that the Lindblad noise operators are at most linear in the quadratures. Under these conditions, an initially Gaussian Wigner function remains
Gaussian throughout the evolution. Therefore, the
dynamics can be described either by a linear Fokker-Planck equation for the Wigner function $W(\bm{x},t)$, or equivalently by closed evolution equations for the first moment $\bm{d}(t)$ and covariance matrix $V(t)$.

In phase space, a time-local Wigner evolution equation
of Fokker-Planck form governs how the quasiprobability
distribution $W(\bm{x},t)$ evolves under drift and
diffusion as \cite{Ernesto2026,Toscano2021,Toscano2022}
\begin{equation}
\begin{aligned}
\frac{\partial W(\bm{x},t)}{\partial t}&=-
\nabla_{\bm{x}}^T \big[\big(A(t)\bm{x}+\bm{u}(t)\big)W(\bm{x},t)\big]\\
&+\frac{1}{2}\nabla_{\bm{x}}^TD(t)\nabla_{\bm{x}} W(\bm{x},t)
\end{aligned}
\end{equation}
where $A(t)$ is the drift matrix describing rotations, squeezing, mode mixing, and attenuation due to photon loss; $\bm{u}(t)$ is the displacement drift vector describing coherent driving or other deterministic shifts of the mean field; and $D(t)$ is the diffusion matrix describing the
effective noise contribution to covariance evolution,
including vacuum noise, thermal noise, and technical
fluctuations within the Gaussian approximation.
In non-Markovian regimes, $D(t)$ need not be positive
semidefinite at every instant. The equation then retains
Fokker-Planck form but need not define a classical
diffusion process. Physicality is imposed through
complete positivity of the accumulated Gaussian channel,
while intermediate propagators need not preserve
positivity for arbitrary input states.

Since the dynamics is Gaussian-preserving, it is fully characterized by the first moment $\bm{d}(t)$ and covariance matrix $V(t)$. Taking the first and second moments of the Fokker-Planck equation gives 
\cite{Ernesto2026,Ma2018,Toscano2021,Toscano2022}
\begin{equation}
\label{eq:first-moment}
\begin{aligned}
\dot{\bm d}(t)
&= \frac{d}{dt}\int \bm x\,W(\bm x,t)\,d\bm x  \\
&= \int \bm x\,\partial_t W(\bm x,t)\,d\bm x \\
&= \int [A(t)\bm x+\bm u(t)]W(\bm x,t)\,d\bm x \\
&= A(t)\bm d(t)+\bm u(t)
\end{aligned}
\end{equation}
and
\begin{equation}
\label{eq:second-moment}
\begin{aligned}
\dot{ V}(t)
&= \frac{d}{dt}\left(\langle \bm x\bm x^T\rangle-\bm d\bm d^T\right) \\
&= A\langle \bm x\bm x^T\rangle+\langle \bm x\bm x^T\rangle A^T
+\bm u\bm d^T+\bm d\bm u^T+D\\
&-\dot{\bm d}\bm d^T-\bm d\dot{\bm d}^{T} \\
&= A\left(\langle \bm x\bm x^T\rangle-\bm d\bm d^T\right)
+\left(\langle \bm x\bm x^T\rangle-\bm d\bm d^T\right)A^T+D \\
&= A(t) V(t)+ V(t)A^T(t)+D(t)
\end{aligned}
\end{equation}
Therefore, the formal solutions of Eqs.\eqref{eq:first-moment} and \eqref{eq:second-moment} are
\begin{align}
\bm{d}(t) &=
X(t)\bm{d}(0)+\int_0^t X(t,s)\bm{u}(s)ds\\
V(t)&=X(t)V(0)X^T(t)+\int_0^t X(t,s)D(s)X^T(t,s)ds
\label{eq:Gaussian-preserving}
\end{align}
where $X(t):=X(t,0)$ describes the deterministic linear transformation and we have
\begin{equation}
\frac{d}{dt}X(t,s)=A(t)X(t,s)
\qquad
X(s,s)=I
\label{eq:X-definition}
\end{equation}
A Gaussian channel is physical and CP
if and only if the following condition is satisfied \cite{Ernesto2026,Mirkhalaf2024}
\begin{equation}
Y(t)+\frac{i}{2}\Omega-\frac{i}{2}X(t)\Omega X^T(t)\geq 0
\label{eq:gaussian-cp}
\end{equation}
where $\Omega$ is the symplectic matrix defined in Eq.\eqref{eq:symplectic} and
\begin{equation}
Y(t)=\int_0^t X(t,s)D(s)X^T(t,s)ds 
\label{eq:Y-definition}
\end{equation}
and so
\begin{equation}
V(t)=X(t)V(0)X^T(t)+Y(t)
\label{eq:gaussian-cp2}
\end{equation}
Here, $Y(t)$ is the accumulated Gaussian noise matrix generated by the diffusion term $D(t)$.
Thus, the pair $X(t),Y(t)$ defines the Gaussian channel acting on the covariance matrix.
While Eq.\eqref{eq:gaussian-cp} ensures CP of the total Gaussian channel from the initial time to $t$, Markovianity in the CP-divisibility sense requires a stronger local condition. Namely, every infinitesimal intermediate map from $t$ to $t+dt$ must itself be CP, which gives
(Appendix \ref{app:Gaussian-propagator})
\begin{equation}
M_{\mathrm{CP}}(t)=D(t)-\frac{i}{2}\left[
A(t)\Omega+\Omega A^T(t)\right]\geq 0
\label{eq:infinitesimal-cp}
\end{equation}
Here $M_{\rm CP}(t)$ is the local Gaussian CP matrix for the infinitesimal intermediate map from $t$ to $t+dt$. Positive semidefiniteness of $M_{\rm CP}(t)$ is equivalent to CP of this infinitesimal Gaussian map. If $M_{\rm CP}(t)$ has a negative
eigenvalue, the intermediate map is not CP, so the dynamics is not CP-divisible. Therefore, violation of Eq.\eqref{eq:infinitesimal-cp}
indicates a breakdown of CP-divisibility and is used here as a witness of non-Markovianity \cite{Rivas2014,Chruscinski2011}. The equivalence between $M_{\mathrm{CP}}(t)\geq0$
throughout an interval and CP-divisibility on that
interval assumes that $A(t)$ is real, $D(t)$ is real
symmetric, and both matrices are continuous, so that
the intermediate Gaussian propagator is well defined.
For such a regular drift matrix
$ \det X(t,s) = \exp\left[\int_s^t
\Tr A(\tau)\,d\tau\right]\neq0
$. The time-local description therefore applies separately
on intervals that exclude singular times. 

Consequently, the Gaussian-preserving formulation makes the Lindblad-like non-Markovian model experimentally and numerically tractable. Instead of
reconstructing the full density operator, the dynamics is reduced to the evolution of the displacement vector $\bm d(t)$ and covariance matrix $V(t)$, which are directly accessible from homodyne measurements in CV photonic
platforms. The reconstructed matrices $A(t)$ and $D(t)$ therefore connect the effective master equation description to measurable phase-space dynamics.

\section{Physical Noise Models}
\label{sec:noise}
\subsection{Physical decomposition of the Gaussian \\noise generator}
\label{sec:hardware_noise}
Here we want to connect the abstract Gaussian generator introduced in the previous section to physical noise mechanisms in CV photonic hardware. This effective generator viewpoint is analogous to microscopic decoherence calculations in other open system settings, where decay rates are derived from the underlying system-environment interaction rather than introduced purely phenomenologically \cite{Carlesso2021,Naeij2022RotationalPhoton,Tirandaz2020Cat}.

In the Gaussian-preserving framework, the first and second moments evolve according to Eqs.\eqref{eq:first-moment} and \eqref{eq:second-moment}, with drift matrix
$A(t)$, diffusion matrix $D(t)$ and displacement drift vector $\bm{u}(t)$. For the noise mechanisms considered below, we decompose the effective Gaussian generator into physically interpretable contributions phenomenologically as
\begin{equation}
A(t)=A_{\mathrm{unitary}}(t)+A_{\mathrm{loss}}(t)
+A_{\mathrm{corr}}(t)+A_{\mathrm{tech}}(t)
\end{equation}
\begin{equation}
D(t)=D_{\mathrm{vac}}(t)+D_{\mathrm{th}}(t)+D_{\mathrm{corr}}(t)+D_{\mathrm{clas}}(t)
\end{equation}
Here, $A_{\mathrm{unitary}}$ describes the intended unitary circuit dynamics, $A_{\mathrm{loss}}$ describes independent photon-loss damping, $A_{\mathrm{corr}}$ describes damping through shared loss channels and $A_{\mathrm{tech}}$ describes deterministic technical deviations from the intended drift. The diffusion terms include $D_{\mathrm{vac}}$, which
describes the vacuum fluctuations introduced when each
optical mode loses photons to an independent vacuum
environment,
additional thermal noise $D_{\mathrm{th}}$, vacuum noise from shared loss channels $D_{\mathrm{corr}}$ and additive classical Gaussian noise $D_{\mathrm{clas}}$. Coherent driving and deterministic displacement offsets are included separately in $\bm{u}(t)$. These expressions organize the selected mechanisms
rather than define a unique microscopic decomposition.
Independent and shared loss contributions are counted
separately, while thermal damping is included in
$A_{\mathrm{loss}}$ and its additional noise in
$D_{\mathrm{th}}$. This separation does not imply that
each contribution can be independently identified from
homodyne data. Even when the probe set is sufficient
to reconstruct the total generator, different sources
may produce identical drift and diffusion matrices.
For example, independent loss and diagonal contributions
from shared loss enter through their sum, while thermal
and additive classical diffusion can be indistinguishable
without additional constraints. Attribution to individual
sources therefore requires independent calibration or
explicit model assumptions.

The following subsections describe the selected physical mechanisms and their contributions to the Gaussian generator, providing explicit expressions where applicable and effective models for technical fluctuations.
\subsubsection{Unitary photonic dynamics}
The coherent dynamics of the CV photonic system is generated by a quadratic
Hamiltonian in Eq.\eqref{eq:Hamiltonian}. Using the canonical commutation relations $[R_j,R_k]=i\Omega_{jk}$,
the Heisenberg equation gives
\begin{equation}
\dot{\bm R}(t) = i[H(t),\bm R(t)] = \Omega G(t)\bm R(t)
\label{eq:unitary_heisenberg_drift}
\end{equation}
Considering Eqs.\eqref{eq:first_moment} and \eqref{eq:first-moment}, the Hamiltonian contribution to the Gaussian drift matrix is
\begin{equation}
\begin{aligned}
&A_{\rm unitary}(t)=\Omega G(t)\\
&\bm{u}_{\rm unitary}(t)=0
\label{eq:unitary_drift}
\end{aligned}
\end{equation}
Under the quadratic Hamiltonian in Eq.\eqref{eq:Hamiltonian}, the covariance
transforms as $V(t)=X(t)V(0)X^T(t)$. Hence,
\begin{equation}
\dot V(t)=A_{\mathrm{unitary}}(t)V(t)
+V(t)A_{\mathrm{unitary}}^T(t)
\end{equation}
which compared with Eq.\eqref{eq:second-moment} gives 
\begin{equation}
D_{\rm unitary}(t)=0 
\label{eq:unitary_diffusion_zero}
\end{equation}
It means that a quadratic Hamiltonian transforms the quadratures without coupling the system to an environment, so it changes the covariance through the drift matrix but introduces no diffusion. 
\subsubsection{Independent photon loss}
For independent photon loss, each mode couples to
its own reservoir, with no cross-mode dissipative
couplings. The Lindblad superoperator for mode $j$ is \cite{Hastrup2021}
\begin{equation}
\mathcal{L}_{\rm loss}^{(j)}\rho
=\kappa_j(t)\mathcal{D}[a_j]\rho
\label{eq:single_loss_revised}
\end{equation}
where $\kappa_j(t)$ is the time-dependent loss rate of mode $j$ and $\mathcal{D}[a_j]$ is the Lindblad dissipator. By independent loss, we mean that the Kossakowski matrix in the basis of mode annihilation operators $a_j$ is diagonal; 
each mode $j$ has its own loss rate
$\kappa_j(t)$ and cross-mode dissipative terms are absent.

For $\kappa_j(t)\geq0$, the Lindblad loss term in Eq.\eqref{eq:single_loss_revised} is equivalently represented in the Markov vacuum-bath limit by the Heisenberg-Langevin equation \cite{Breuer2007,Gardiner1985}
\begin{equation}
\dot a_j(t)=-\frac{\kappa_j(t)}{2}a_j(t)
+\sqrt{\kappa_j(t)}\,a_{j,\mathrm{in}}(t)
\label{eq:langevin_loss}
\end{equation}
where $a_{j,\mathrm{in}}(t)$ denotes the vacuum input field of the reservoir coupled to mode $j$. Using $a_j=\frac{1}{\sqrt{2}} (q_j+ip_j)$ and defining the single-mode quadrature vectors $\bm R_j=(q_j,p_j)^T$, Eq.\eqref{eq:langevin_loss} gives
\begin{equation}
\dot {\bm{R}}_j(t)=-\frac{\kappa_j(t)}{2}\bm{R}_j(t)
+\sqrt{\kappa_j(t)}\bm{R}_{j,\mathrm{in}}(t)
\end{equation}
The vacuum input satisfies
\begin{equation}
\langle a_{j,\rm in}(t)a_{j,\rm in}^\dagger(t')\rangle=\delta(t-t'),
\quad
\langle a_{j,\rm in}^\dagger(t)a_{j,\rm in}(t')\rangle=0 
\end{equation}
and using Eq.\eqref{eq:quadrature}, symmetrized correlations can be obtained
\begin{equation}
\frac{1}{2}\left\langle\left\{R_{j,\mathrm{in},\mu}(t),R_{j,\mathrm{in},\nu}(t')\right\}\right\rangle=
\frac{1}{2}\delta_{\mu\nu}\delta(t-t') 
\end{equation}
which is consistent with the convention $[q_j,p_j]=i$ and vacuum
covariance $I_2/2$.
Here $\mu,\nu\in\{1,2\}$ label the components of the reservoir input quadrature vector, corresponding to $R_{j,\mathrm{in},1}=q_{j,\mathrm{in}}$ and
$R_{j,\mathrm{in},2}=p_{j,\mathrm{in}}$.

Over a small time interval $dt$, a pure-loss channel with rate $\kappa_j(t)$ has transmissivity \cite{Weedbrook2012}
\begin{equation}
\eta_j(t)=1-\kappa_j(t)dt+O(dt^2)
\end{equation}
where $\eta_j(t)$ quantifies the fraction of the input
optical intensity transmitted through the loss channel, with $\eta_j=1$ corresponding to no loss and $\eta_j=0$ to complete attenuation.
Moreover, a pure-loss Gaussian channel describes attenuation
with no added noise beyond the vacuum fluctuations required
by quantum mechanics. It can be represented as a beam-splitter
interaction between the system mode and an environmental
vacuum mode.
Accordingly, the quadratures transform as \cite{Weedbrook2012, Gardiner2004}
\begin{equation}
 \bm R_j(t+dt) =
\sqrt{\eta_j(t)} \bm R_j(t) +
\sqrt{1-\eta_j(t)} \bm R_{j,\rm vac}
\end{equation}
where $\bm R_{j,\mathrm{vac}}$ is an independent vacuum quadrature vector. Substituting $\eta_j(t)$, taking the expectation value based on Eq.\eqref{eq:first_moment} and
using $\langle \bm R_{j,\rm vac}\rangle=0$, the first moment evolves as
\begin{equation}
d_j(t+dt)=\left(1-\frac{\kappa_j(t)}{2}dt\right)d_j(t)+O(dt^2)
\end{equation}
and the covariance matrix in Eq.\eqref{eq:covariance} evolves as
\begin{equation}
V_j(t+dt)=\left(1-\kappa_j(t)dt\right)V_j(t)
+\frac{\kappa_j(t)}{2}I_2 dt+
O(dt^2)
\end{equation}
Subtracting the moments at time $t$ from both sides, dividing by $dt$, and taking the limit $dt\to0$ yields
\begin{equation}
\dot d_j(t)=-\frac{\kappa_j(t)}{2}d_j(t),
\qquad
 \dot V_j(t)=-\kappa_j(t)V_j(t) +\frac{\kappa_j(t)}{2}I_2 
\end{equation}
Therefore, comparison with the Gaussian moment equation in Eqs.\eqref{eq:first-moment} and \eqref {eq:second-moment} identifies the independent photon-loss contribution as
\begin{equation}
A_{\mathrm{loss},j}(t)=-\frac{\kappa_j(t)}{2}I_2
\qquad
D_{\mathrm{vac},j}(t)=
\frac{\kappa_j(t)}{2}I_2 
\label{photon-loss}
\end{equation}
For independent loss in an $m$-mode system, these single-mode blocks give
\begin{equation}
\begin{aligned}
&K_{\kappa}(t)=\bigoplus_{j=1}^{m}\kappa_j(t)I_2\\
&A_{\rm loss}(t)=-\frac{1}{2}K_{\kappa}(t)\\
&D_{\rm vac}(t)=\frac{1}{2}K_{\kappa}(t)
\end{aligned}
\end{equation}
The vacuum Langevin and infinitesimal beam-splitter
derivations above assume $\kappa_j(t)\geq0$.
For negative effective rates, the same drift and
diffusion relations follow instead by differentiating
the accumulated pure-loss channel, provided its
transmissivity remains between zero and one.
Such rates describe a temporary increase in the
accumulated transmissivity, rather than a loss step
with a fresh, independent vacuum input.

\subsubsection{Thermal-loss channel}
Thermal excitation describes the situation in which the effective environment is not in the vacuum state, but contains a finite average number of photons. In this case, the photonic mode is not only damped by photon loss, but can also absorb excitations from the bath. Therefore, the master equation contains two contributions: a loss term proportional to $\mathcal{D}[a_j]\rho$, and a heating term proportional to $\mathcal{D}[a_j^\dagger]\rho$. The strength of the heating contribution is controlled by the effective thermal occupation $\bar n_j(t)$ of the environment coupled to mode $j$. The thermal-loss contribution to the Lindblad master
equation can therefore be written as \cite{Gardiner2004}
\begin{equation}
\mathcal{L}_{\rm th}^{(j)}\rho=
\kappa_j(t)\big(\bar n_j(t)+1\big)\mathcal{D}[a_j]\rho
+
\kappa_j(t)\bar n_j(t)\mathcal{D}[a_j^\dagger]\rho
    \label{eq:thermal_master_revised}
\end{equation}
where $\mathcal{L}_{\rm th}^{(j)}$ means the thermal Lindblad superoperator for mode j.

To obtain the additional thermal diffusion, we separate the vacuum-loss contribution in Eq.\eqref {eq:single_loss_revised} from the
thermal-loss generator in Eq.\eqref{eq:thermal_master_revised}. The remaining term is $\kappa_j(t)\bar n_j(t)
\bigl(\mathcal D[a_j]+\mathcal D[a_j^\dagger]\bigr)\rho$. Its loss and gain contributions produce opposite drift terms which cancel, while each contributes
$\kappa_j(t)\bar n_j(t)I_2/2$ to diffusion.
Comparison with Eq.\eqref {eq:second-moment} therefore gives the additional single-mode diffusion $\kappa_j(t)\bar n_j(t)I_2$.
Collecting these blocks for independent modes yields
\begin{equation}
D_{\rm th}(t)=\bigoplus_{j=1}^{m}\kappa_j(t)\bar n_j(t) I_2 
\label{eq:D_th_revised}
\end{equation}
Here $D_{\rm th}(t)$ denotes the additional thermal diffusion beyond the
vacuum diffusion $D_{\rm vac}(t)$ already associated with photon loss in Eq.\eqref {photon-loss}.
Therefore, the total diffusion associated with a thermal loss channel is
\begin{equation}
D_{\rm loss+th}(t)=D_{\rm vac}(t)+D_{\rm th}(t)
=\bigoplus_{j=1}^{m}\kappa_j(t)\left(\bar n_j(t)+\frac{1}{2}\right)I_2 
\label{eq:D_total_thermal_revised}
\end{equation}
Moreover, the drift matrix $A_{\rm loss}$ remains the same as in Eq.\eqref {photon-loss}, since the additional thermal loss and excitation terms make equal and opposite contributions to the drift.

In CV photonic systems, the thermal occupation $\bar n_j(t)$ is often negligible because the environment is well approximated by vacuum. We keep this term in the model for generality, especially for microwave, optomechanical, or hybrid CV platforms where thermal photons may contribute appreciably to the noise.

\subsubsection{Correlated photon loss}
In CV photonic systems, loss may not be completely independent between modes.  For example, neighboring waveguides, common scattering
regions or shared spectral modes can create correlated dissipation.  In this case, the
standard correlated photon-loss Lindblad form is
\begin{equation}
\mathcal{L}_{\rm corr}\rho=\sum_{j,k=1}^{m}\Gamma_{jk}(t)\left(a_j\rho a_k^\dagger-\frac{1}{2}\{a_k^\dagger a_j,\rho\}\right)
\label{eq:corr_loss_revised}
\end{equation}
where $\Gamma(t)=\Gamma^\dagger(t)$ is the correlated-loss matrix. The diagonal terms
$\Gamma_{jj}(t)$ describe ordinary single-mode loss, while the off-diagonal terms $\Gamma_{jk}(t)$ describe correlated loss between
modes $j$ and $k$.
For a Markovian correlated loss channel, the matrix must satisfy \cite{Breuer2007,Breuer2009,Rivas2014,Schlosshauer2007}.
\begin{equation}
\Gamma(t)\geq 0
\label{eq:Gamma_positive_revised}
\end{equation}
In a time-local description, $\Gamma(t)$ may
temporarily have negative eigenvalues. For the
isolated correlated-loss model, this implies
CP-indivisibility, although the accumulated channel
may remain CP. When other loss or diffusion
mechanisms are present, their contributions may
compensate these negative directions. CP-divisibility
must then be assessed using $M_{\mathrm{CP}}(t)$
constructed from the total drift and diffusion
matrices, rather than from $\Gamma(t)$ alone.

For simplicity, we take the correlated-loss matrix
$\Gamma(t)$ to be real and symmetric. It then produces identical dissipative coupling among the $q$ quadratures and among the $p$ quadratures, without mixing the two. With the interleaved quadrature ordering defined in
Eq.\eqref{eq:quadrature2}, the corresponding matrix in quadrature space is
\begin{equation}
B_\Gamma(t)=\Gamma(t)\otimes I_2
\label{eq:BGamma_revised}
\end{equation}
To determine how the correlated-loss matrix enters the Gaussian drift, we evaluate the evolution of the mode amplitudes. From Eq.\eqref{eq:corr_loss_revised}
\begin{equation}
\frac{d}{dt}\langle a_n\rangle
= \operatorname{Tr}\!\left[a_n\mathcal{L}_{\rm corr}\rho\right]
= -\frac{1}{2}\sum_{j=1}^{m}
\Gamma_{jn}(t)\langle a_j\rangle
\label{eq:quadraure_loss}
\end{equation}
For a general Hermitian correlated-loss matrix, this
corresponds to the amplitude drift
$-\Gamma^{T}(t)/2$. Since we restrict $\Gamma(t)$
to be real and symmetric, $\Gamma^{T}(t)=\Gamma(t)$,
and the quadrature-space representation remains unchanged.
Thus, the diagonal elements of $\Gamma(t)$ describe ordinary modal attenuation, while the off-diagonal elements generate dissipative
coupling between different mode amplitudes. This relation provides the connection between the correlated-loss master equation and the
phase-space drift matrix.
Using the quadrature definitions in Eq.\eqref{eq:quadrature} and the
quadrature-space loss matrix $B_\Gamma(t)$, Eq.\eqref{eq:quadraure_loss} becomes
\begin{equation}
\dot{\mathbf d}(t) =-\frac{1}{2}B_\Gamma(t)\mathbf d(t)
\end{equation}
Comparison with Eq.\eqref{eq:first-moment} then shows that the
correlated-loss contribution to the drift matrix is
\begin{equation}
A_{\rm corr}(t)=-\frac{1}{2}B_\Gamma(t)
\label{eq:corr_AD_revised}
\end{equation}
Since the loss generator preserves the vacuum covariance, Eq.\eqref{eq:second-moment} requires
\begin{equation}
0=(A_{\mathrm{corr}}+A_{\mathrm{corr}}^T)/2
+D_{\mathrm{corr}}
\end{equation}
Substituting Eq.\eqref{eq:corr_AD_revised} and using $B_\Gamma^T=B_\Gamma$ gives
\begin{equation}
D_{\rm corr}(t)=\frac{1}{2}B_\Gamma(t)
\label{eq:D_corr}
\end{equation}
Eq.\eqref{eq:corr_AD_revised} is the multimode analogue of the
single-mode photon-loss relations in Eq.\eqref{photon-loss}. In this case,
the scalar loss rate $\kappa_j(t)$ is replaced by the quadrature-space
loss matrix $B_\Gamma(t)$. The drift term $A_{\rm corr}(t)$ describes the contraction of the multimode phase-space distribution due to correlated
loss, while the diffusion term $D_{\rm corr}(t)$ represents the accompanying vacuum noise required by the loss channel. The factor $1/2$ in $A_{\mathrm{corr}}$ reflects amplitude
decay at half the photon-number decay rate.
The corresponding $D_{\mathrm{corr}}$ follows from the same loss generator and ensures that the vacuum
covariance remains unchanged.

\subsubsection{Classical technical fluctuations}
\label{sec:technical_fluctuations}
In addition to quantum loss and thermal excitation, CV photonic systems are also affected by classical technical noise. Important examples include pump-amplitude fluctuations, pump-phase noise, slow thermal drift, imperfect phase stabilization, and fluctuations of external control fields. These noise sources are classical in the sense that they can often be modeled as stochastic variations of system parameters rather than as independent quantum bath modes.
For example, pump-amplitude fluctuations can change the effective squeezing strength, while phase fluctuations can perturb phase shifters, interferometers or homodyne reference phases. 

To describe technical fluctuations within the Gaussian
framework, we consider additive noise represented by a real,
zero-mean Gaussian stochastic vector $\boldsymbol{\xi}(t)$,
coupled to the quadratures through a real matrix $B_{\xi}(t)$.
This assumption applies directly to Gaussian random
displacements, but does not generally describe fluctuations
of phase or squeezing strength, which act multiplicatively
and require an additional linearization or effective Gaussian
approximation. A nonzero mean can be included in the
displacement drift $\boldsymbol{u}(t)$. In the white-noise
limit, valid when the noise correlation time is much shorter
than the relevant system evolution timescales, we have
\cite{Breuer2007,Gardiner2004}
\begin{equation}
\mathbb{E}[\boldsymbol{\xi}(t)\boldsymbol{\xi}^{T}(s)] =Q_{\xi}(t)\delta(t-s),\quad Q_{\xi}(t)\geq0
\label{eq:white_noise}
\end{equation}
specifies the noise strengths and correlations. Here $\mathbb{E}$ denotes an ensemble average over noise realizations. Since the additive noise has zero mean, it contributes neither to $A(t)$ nor to $\bm{u}(t)$ in the first-moment equation. It nevertheless changes the covariance through the additional diffusion matrix
\begin{equation}
D_{\mathrm{clas}}(t)=B_{\xi}(t)Q_{\xi}(t)B_{\xi}^T(t)\geq0
\end{equation}

More generally, for noise with finite temporal correlations, the white-noise relation in Eq.\eqref{eq:white_noise} is replaced by \cite{Gardiner2004}
\begin{equation}
\mathbb{E}[\xi_\mu(t)\xi_\nu(s)] =C_{\mu\nu}(t,s)
\end{equation}
where $\mu$ and $\nu$ label the classical noise components. Eq.\eqref{eq:white_noise} is recovered when $C_{\mu\nu}(t,s)=[Q_\xi(t)]_{\mu\nu}\delta(t-s)$.  When the correlation time is much shorter than the system evolution timescale, the white-noise approximation above is appropriate. Otherwise, the effective diffusion depends on the noise correlations and the intervening system evolution, and cannot generally be expressed by the same instantaneous formula. This description includes finite-correlation-time processes
whose reduced dynamics preserves Gaussian states and admits
a well-defined time-local generator; it does not cover
arbitrary temporally correlated noise. Within this class,
temporal correlations may lead to violations of
CP-divisibility, but a finite correlation time alone does
not guarantee such a violation.

In the time-local Gaussian model, we do not explicitly
reconstruct the underlying classical stochastic process.
Instead, zero-mean additive fluctuations are represented
through the effective diffusion term $D_{\mathrm{clas}}(t)$
in the covariance dynamics of Eq.\eqref{eq:second-moment}.
For temporally correlated fluctuations, $D_{\mathrm{clas}}(t)$ should be interpreted as an effective time-local coefficient rather
than as the microscopic two-time correlation function. Such a coefficient may be time dependent while the dynamics remains CP-divisible. Non-Markovianity is identified only when the
reconstructed time-local generator violates the infinitesimal Gaussian CP condition while the accumulated finite-time channel remains CP.

\subsection{Time-correlated noise kernels and effective rates}
\label{subsec:time_correlated_noise}
The noise mechanisms discussed above were introduced through time-local coefficients in the Gaussian generator. However, in realistic CV photonic hardware these coefficients can originate from underlying stochastic processes
with finite temporal correlations. The goal of this section is to distinguish the microscopic two-time noise kernels from the effective local rates that enter the reconstructed matrices $A(t)$ and $D(t)$.

Temporal correlations can enter any of the hardware noise coefficients introduced above, including photon loss, thermal occupation, correlated dissipation, and classical control fluctuations. We denote these noise parameters collectively by
\begin{equation}
\boldsymbol{\zeta}(t) = \{\kappa_j(t),\,\bar n_j(t),\,\Gamma_{jk}(t),\, \xi_{\mu}(t)\}
\label{eq:noise_parameter_vector}
\end{equation}
where $\kappa_j(t)$ is the effective photon-loss rate of mode $j$ and $\bar n_j(t)$ is the effective occupation of its thermal environment. The entries $\Gamma_{jk}(t)$ describe correlated loss between modes $j$ and $k$, while
$\xi_\mu(t)$ is the $\mu$th component of the classical stochastic noise vector introduced in Eq.\eqref{eq:white_noise}. When
matrix-valued quantities such as $\Gamma(t)$ are included in $\boldsymbol{\zeta}(t)$, their independent entries are collected into a single parameter vector. The microscopic parameter fluctuations are
characterized by the two-time covariance matrix \cite{Breuer2007}
\begin{equation}
C_{\zeta}(t,s)=\mathbb E[\delta\boldsymbol{\zeta}(t)
\delta\boldsymbol{\zeta}^{T}(s)],
\qquad
\delta\boldsymbol{\zeta}(t)=\boldsymbol{\zeta}(t)-\mathbb E[\boldsymbol{\zeta}(t)] 
\label{eq:noise_parameter_correlations}
\end{equation}
Here $\boldsymbol{\zeta}(t)$ is understood as a column vector obtained by collecting the
microscopic stochastic noise parameters, with matrix-valued quantities such as
$\Gamma_{jk}(t)$ vectorized when necessary.

For a stationary noise source, the correlation depends only on the time delay $\tau=t-s$, so that $C_{\zeta}(t,s)=C_{\zeta}(\tau)$. The same
information can then be represented by the noise spectral-density matrix \cite{Breuer2007,Rivas2014}
\begin{equation}
C_{\zeta}(\tau)=\int \frac{d\omega}{2\pi}\,
S_{\zeta}(\omega)e^{-i\omega\tau},
\qquad
\tau=t-s 
\label{eq:noise_spectral_density}
\end{equation}
where $\omega$ is the angular-frequency variable, and $S_{\zeta}(\omega)$ describes the frequency content of the correlated noise. Structured features
of $S_{\zeta}(\omega)$ correspond to finite bandwidth reservoirs, resonator memory, delay-line feedback, or slowly relaxing technical noise.

The effect of temporal correlations depends on the physical sector. Photon loss and correlated dissipation modify both the drift and the
associated vacuum diffusion, whereas thermal occupation and classical fluctuations primarily contribute additional diffusion. For example, using the same coupling matrix $B_\xi(t)$ introduced above,
classical fluctuations with finite temporal correlations give the phase-space noise kernel \cite{Gardiner2004,Breuer2007}
\begin{equation}
\mathcal N_{\mathrm{cl}}(t,s)= {B}_\xi(t)\, C_{\xi}(t,s)\, {B}^{T}_\xi(s)
\label{eq:classical_noise_kernel}
\end{equation}
with $C_{\xi}(t,s)=\mathbb E[\boldsymbol{\xi}(t)\boldsymbol{\xi}^{T}(s)]$. This relation follows by defining the effective phase-space noise force as
$\mathbf{f}(t)=B_{\xi}(t)\boldsymbol{\xi}(t)$ and evaluating its two-time covariance
$\mathbb{E}[\mathbf{f}(t)\mathbf{f}^{T}(s)]$.

In the white-noise limit, this kernel becomes local in time and reduces to an ordinary diffusion term. For finite-bandwidth fluctuations, however, the noise is more naturally described by a memory kernel; in the time-local reconstruction this appears as a time-dependent effective diffusion.
Thus, the quantities $\kappa_j(t)$,
$\Gamma_{jk}(t)$, and $D_{\mathrm{clas}}(t)$ should be interpreted as effective local coefficients associated with underlying two-time noise
correlations, not as the microscopic memory kernels themselves. This distinction is important because time dependence alone does not imply
non-Markovianity. A time-dependent Gaussian generator can still be CP-divisible while non-Markovian behavior requires violation of the
infinitesimal CP-divisibility condition.

\subsection{Markovian and non-Markovian regimes}
\label{subsec:markovian_nonmarkovian_regimes}
The physical noise models above can generate either Markovian or non-Markovian Gaussian dynamics, depending on the local divisibility
properties of the effective generator. Time dependence of the coefficients
is not sufficient to identify memory: a process with time-dependent loss,
thermal occupation, dephasing, correlated dissipation, or classical
diffusion may still be CP-divisible.

Using the infinitesimal Gaussian CP matrix $M_{\rm CP}(t)$ defined in
Eq.\eqref{eq:infinitesimal-cp}, we classify the dynamics in the divisibility sense as
\begin{equation}
\begin{cases}
\lambda_{\min}[M_{\rm CP}(t)] \geq 0 \quad \forall t,
& \text{CP-divisible regime} \\
\lambda_{\min}[M_{\rm CP}(t)] < 0 \quad \text{for some } t,
& \text{non-CP-divisible regime}
\end{cases}
\label{eq:regime_classification}
\end{equation}
In the second case, the negative eigenvalue indicates that the infinitesimal
intermediate map from $t$ to $t+dt$ is not CP. This is the Gaussian CP-divisibility witness used here to identify non-Markovianity in the
divisibility sense \cite{Breuer2009,Rivas2014}.
This local violation must be distinguished from loss of physicality of the full evolution. The finite-time Gaussian channel must still satisfy the
CP condition in Eq.\eqref{eq:gaussian-cp}. Within the present framework, the dynamics is
therefore identified as non-Markovian only when
$M_{\mathrm{CP}}(t)$ has a negative eigenvalue over a finite time interval while the finite-time Gaussian
channel remains CP. Time-dependent coefficients
or finite temporal correlations alone are not sufficient to establish non-Markovianity in the CP-divisibility sense.

\subsection{Markovian embedding of non-Markovian Gaussian noise}
\label{subsec:pseudomode_embedding}
The time-local Lindblad-like generator gives an effective description of the
physical photonic modes alone. When some effective coefficients become temporarily negative, the reduced generator is not a standard Lindblad
generator at that instant. This does not imply that the full dynamics is unphysical. A standard interpretation is that the observed modes are coupled
to unobserved memory-bearing degrees of freedom, such as resonator modes,
delay-line modes, hidden spectral modes, or structured reservoirs
\cite{Garraway1997,Dalton2001,Pleasance2020,
Pleasance2021}.

Let $a_j$ denote the physical photonic modes and $b_\ell$ auxiliary bosonic
memory modes. A Gaussian-preserving embedded model may be written as \cite{Pleasance2020}
\begin{equation}
H_{\rm emb}=H_{\rm ps}+\sum_{\ell}\omega_{b,\ell} b_\ell^\dagger b_\ell+\sum_{j,\ell}\left(
g_{j\ell} a_j^\dagger b_\ell+
g_{j\ell}^{*} a_j b_\ell^\dagger
\right)
\label{eq:embedded_hamiltonian}
\end{equation}
where $H_{\rm ps}$ is the quadratic Hamiltonian of the photonic system and
$g_{j\ell}$ controls the coupling to the memory modes and $\omega_{b,\ell}$ is the resonance frequency of the $\ell$-th auxiliary memory mode. The auxiliary modes $b_\ell$ act as memory-bearing degrees of freedom, while their coupling to external Markovian reservoirs is represented
through the damping terms introduced below. For each input probe, the initial joint state is
$\varrho(0)=\rho_S(0)\otimes\rho_M$, where
$\rho_S(0)$ is the photonic input state and $\rho_M$
is a fixed Gaussian memory state, identical for all
probes.
The pseudomodes couple to independent Markovian
reservoirs, initially uncorrelated with the system
and memory. Under the Gaussian-preserving enlarged
dynamics, these assumptions define a Gaussian CPTP
reduced channel for arbitrary system inputs.

The joint density operator $\varrho(t)$ of the physical photonic modes and the auxiliary memory modes evolves
according to the Markovian Lindblad equation
\begin{equation}
\frac{d\varrho}{dt}=-i[H_{\rm emb},\varrho]
+
\sum_{\ell}\gamma_\ell \mathcal D[b_\ell]\varrho 
\label{eq:embedded_lindblad}
\end{equation}
with positive pseudomode decay rates $\gamma_\ell$. Thus, the external environment does not appear as an explicit dynamical
mode in the enlarged Hamiltonian; instead, its influence is encoded in
the positive pseudomode decay rates $\gamma_\ell$ and the associated
Lindblad dissipators $\mathcal{D}[b_\ell]$. The physical photonic
state is obtained after tracing out the memory modes
\begin{equation}
\rho(t)={\Tr}_{b}\{\varrho(t)\}
\label{eq:reduced_photonic_state}
\end{equation}
Formally solving the damped equation of motion for the auxiliary modes $b_\ell$ and substituting the result into the equations for the photonic modes produces a time-nonlocal term proportional to
$\sum_k\int_0^t K_{jk}(t-s)a_k(s)\,ds$, together with a reservoir-noise term. For the passive, number-preserving coupling in Eq.\eqref{eq:embedded_hamiltonian}, eliminating the damped pseudomodes gives the memory kernel
\cite{Garraway1997,Pleasance2021}
\begin{equation}
K_{jk}(t-s)=\sum_{\ell}g_{j\ell}g_{k\ell}^{*}e^{-(i\omega_{b,\ell}+\gamma_\ell/2)(t-s)}\Theta_H(t-s)
\label{eq:pseudomode_memory_kernel}
\end{equation}
Here, $\gamma_\ell$ is the damping rate of pseudomode
$\ell$ and the Heaviside function $\Theta_H(t-s)$
enforces causality, while the exponential describes
oscillation at frequency $\omega_{b,\ell}$ and
amplitude decay as $e^{-\gamma_\ell(t-s)/2}$,
giving an amplitude-memory decay time
$2/\gamma_\ell$ for $\gamma_\ell>0$.
The diagonal elements $K_{jj}(t-s)$ describe memory in individual loss channels, while
the off-diagonal elements $K_{jk}(t-s)$ describe correlated dissipative memory shared
between modes. Thus, a reduced photonic generator may become temporarily non-CP-divisible
even though the joint system evolves under a CP
Markovian Lindblad equation.

The pseudomode embedding plays three roles in the present framework. First, it provides a physical mechanism for the time-dependent and possibly negative coefficients that appear in the reduced Lindblad-like generator. Second, it
shows how reduced non-CP-divisible dynamics can arise from a CP Markovian evolution on an enlarged system. 
Third, it connects phenomenological parameters such as
$\gamma_\ell$, $\omega_{b,\ell}$, and $g_{j\ell}$
to hardware-level quantities such as pseudomode lifetime,
detuning, and coupling strength.
Therefore, pseudomode embedding provides a bridge between the phenomenological time-local tomography model and a microscopic hardware model of non-Markovian
photonic noise.

The theoretical framework developed in the preceding sections motivates the
reconstruction strategy introduced below. Rather than directly identifying all
microscopic reservoirs, hidden memory modes, or two-time noise kernels, we
reconstruct their effect through the time-local Gaussian generator
$A(t)$, $D(t)$, and $\bm u(t)$.

\section{Tomographic Reconstruction Method}
\label{sec:tomogtaphy}
Quantum tomography provides a framework for reconstructing unknown quantum states, processes, or effective dynamical models from measurement outcomes \cite{DAriano2024,Merkel2013}. Here, we develop a theoretical and simulation-based tomography framework for CV photonic systems whose central objective is to move beyond fixed-time Gaussian state reconstruction and instead learn an effective time-local description of the noise dynamics. The reconstruction procedure combines Gaussian moment propagation, homodyne probability distributions, maximum-likelihood estimation, physicality constraints, and statistical model comparison. This makes it possible to determine whether the observed or simulated homodyne data are consistent with CP-divisible Markovian Gaussian noise, or whether they require a non-Markovian time-local Gaussian description.

\subsection{Tomographic reconstruction of non-Markovian Gaussian noise dynamics}
The unknown noise dynamics is modeled by an effective time-local Gaussian generator. In the Gaussian-preserving regime, this generator is fully characterized at the level of first and second moments in Eq.\eqref{eq:Gaussian-preserving} by the drift matrix $A(t)$, the diffusion matrix $D(t)$, and the displacement drift $\bm{u}(t)$. When the coherent part of the system is also unknown, the Hamiltonian matrix $G(t)$ may be included through the relation $A_{\rm unitary}(t)=\Omega G(t)$. These quantities determine the evolution of the displacement vector and covariance matrix, which in turn determine the predicted homodyne statistics at each measurement time. The reconstruction problem is therefore formulated as a statistical inverse problem: choose the generator parameters so that the predicted quadrature distributions best match the measured or synthetic time-resolved homodyne data while preserving the physicality of the total Gaussian channel.

Suppose one prepares a set of input Gaussian states
\begin{equation}
\mathcal{S}_{\mathrm{in}}=
\{\rho_\alpha(0)\}_{\alpha=1}^{N_{\mathrm{in}}}
\end{equation}
including coherent states, squeezed states, displaced squeezed states, and two-mode squeezed states. At measurement times
\begin{equation}
\mathcal{T}=\{t_1,t_2,\ldots,t_K\}
\end{equation}
one performs homodyne measurements of quadratures
\begin{equation}
X_{j,\theta}=q_j\cos\theta+p_j\sin\theta
=\bm e_{j,\theta}^{T}\bm R
\end{equation}
where $j$ labels the measured mode, $\theta$ is the homodyne measurement phase, and $\bm e_{j,\theta}$ is the corresponding unit vector in phase space. A homodyne measurement of $X_{j,\theta}$ yields a real-valued outcome
$x$. For a Gaussian state, this outcome $x$ is normally distributed \cite{Vasile2011}
\begin{equation}
p(x|j,\theta,t,\alpha;\Theta)
=\frac{1}{\sqrt{2\pi\sigma^2_{j,\theta,\alpha}(t)}}\exp\left[
-\frac{\left(x-\mu_{j,\theta,\alpha}(t)\right)^2}{2\sigma^2_{j,\theta,\alpha}(t)}
\right]
\label{eq:homodyne-probabilities}
\end{equation}
where
\begin{align}
\mu_{j,\theta,\alpha}(t)&=
\bm e_{j,\theta}^{T}\bm d_\alpha(t)
\label{eq:mean}\\
\sigma^2_{j,\theta,\alpha}(t)
&=\bm e_{j,\theta}^{T}V_\alpha(t)\bm e_{j,\theta}
\label{eq:variance}
\end{align}
Here, $\Theta$ denotes the set of unknown model parameters to be estimated, including the parameters that define $A(t)$, $D(t)$, $\bm{u}(t)$ and possibly $G(t)$. The vector $\bm e_\theta$ is the corresponding unit vector in phase space; for a single mode $\bm e_\theta=(\cos\theta,\sin\theta)^T$. 
For an $m$-mode system, $\bm e_\theta\in\mathbb{R}^{2m}$ selects the measured mode and quadrature direction. The vector $\bm d_\alpha(t)$ is the first-moment displacement vector of the evolved state that originated from the input probe state $\rho_\alpha(0)$ and $V_\alpha(t)$ is its covariance matrix. The label $\alpha$ identifies which input probe state was prepared. The quantity $\mu_{j,\theta,\alpha}(t)$ is the predicted mean value of the measured quadrature and $\sigma^2_{j,\theta,\alpha}(t)$ is the corresponding quadrature variance.

For multimode reconstruction, we also record joint homodyne
outcomes of commuting quadratures on distinct modes.
For a setting represented by a matrix $H$ whose rows select
these quadratures, the joint outcome is Gaussian with mean
$H \bm d_\alpha(t)$ and covariance $HV_\alpha(t)H^T$. Measuring
each mode pair at phase combinations
$(0,0)$, $(0,\pi/2)$, $(\pi/2,0)$, and $(\pi/2,\pi/2)$
determines its intermode covariance block. Accordingly,
the likelihood below uses joint Gaussian densities for
multimode measurements, retaining correlations within each
shot and assuming independence only between shots.
The scalar formulas apply to the single-mode benchmarks;
identification of the full generator additionally requires
sufficiently informative probes and a specified temporal
parameterization.

Let the homodyne data set, either experimentally collected or synthetically generated in numerical simulations, be
\begin{equation}
\mathcal{D} =\left\{x_n,j_n,\theta_n,t_n,\alpha_n
\right\}_{n=1}^{N}
 \label{eq:data-set}
\end{equation}
Here, $n$ labels the individual homodyne samples in the data set. For a fixed choice of unknown model parameters $\Theta$, the Gaussian moment
relations in Eqs.\eqref{eq:first-moment} and \eqref{eq:second-moment} are solved for every input state $\rho_{\alpha}(0)$. This gives the predicted mean vector $\bm d_{\alpha}(t)$ and covariance matrix $V_{\alpha}(t)$ at all measurement times.

We adopt a maximum-likelihood approach to estimate the unknown generator parameters $\Theta$ from homodyne data \cite{James2001,Banaszek1999,Fiurasek2001,Sacchi2001,Lvovsky2009}. For each trial value of $\Theta$, the Gaussian moment equations determine the predicted mean and variance of every measured quadrature, and hence the probability density for each homodyne outcome. The likelihood is defined as the joint probability density of obtaining the observed data set under these model predictions. 
Assuming statistically independent homodyne shots conditioned on $\Theta$, the total likelihood is therefore given by the product of the single-shot probability densities
\begin{equation}
\mathfrak{J}(\Theta|\mathcal{D})=\prod_{n=1}^{N}
p(x_n|j_n,\theta_n,t_n,\alpha_n;\Theta)
\label{eq:likelihood-product}
\end{equation}

For numerical optimization, we work with the negative log-likelihood. Equivalently, one minimizes the negative log-likelihood
\begin{equation}
\mathcal{J}(\Theta)=-\sum_{n=1}^{N}\log
p(x_n|j_n, \theta_n,t_n,\alpha_n;\Theta)
\label{eq:negative-log-likelihood}
\end{equation}
The maximum-likelihood estimator is therefore
\begin{equation}
\Theta_{\mathrm{ML}}
=\arg\min_{\Theta}
\mathcal{J}(\Theta)
\label{eq:ml-estimator}
\end{equation}
where $\Theta_{\mathrm{ML}}$ is the set of model parameters that gives the best fit to the homodyne data. However, reconstructing a time-dependent generator from finite data is an ill-conditioned inverse problem. So regularization can be used to obtain stable estimates
\cite{Clason2021}. Without additional constraints,
many rapidly oscillating functions $A(t)$ and $D(t)$
may fit the same noisy data. Therefore, we introduce
a regularized estimator for general generator
reconstruction \cite{Banaszek1999,Clason2021}
\begin{equation}
\Theta_{\mathrm{reg}}=\arg\min_{\Theta}\left[\mathcal{J}(\Theta)+\lambda_{\mathrm{s}}R_{\mathrm{smooth}}(\Theta)+\lambda_{\mathrm{p}}R_{\mathrm{phys}}(\Theta)\right]
\label{eq:regularized-estimator}
\end{equation}
where the first term, $\mathcal{J}(\Theta)$ is the negative log-likelihood and measures how well the model explains the homodyne data; the term $R_{\rm smooth}(\Theta)$ penalizes unphysical rapid temporal variations of the reconstructed generator, while $R_{\rm phys}(\Theta)$ penalizes violations of CP of the total Gaussian channel. The non-negative hyperparameters $\lambda_s$ and $\lambda_p$ control the relative importance of smoothness and physicality compared with data fitting. Thus, Eq.\eqref{eq:regularized-estimator} estimates the most likely generator that is simultaneously consistent with the measured data, temporally regular, and physically valid.

We use a first-order Tikhonov regularizer
\cite{Clason2021} which penalizes the squared temporal
derivatives of the reconstructed generator coefficients
\begin{equation}
R_{\mathrm{smooth}}(\Theta) =\int_{0}^{T}\left[\left\|\dot{A}(t)\right\|_{F}^{2}
+\left\|\dot{D}(t)\right\|_{F}^{2}+\left\|\dot{\bm{u}}(t)\right\|_{2}^{2}\right]dt
\label{eq:smoothness-regularizer}
\end{equation}
Here $\|\cdot\|_{F}$
denotes the Frobenius norm, while $\|\cdot\|_2$ denotes the Euclidean norm for vectors. 
This term encodes the physical assumption that hardware noise parameters do not vary arbitrarily fast compared with the experimental sampling time.

The second regularizer penalizes departures from physicality of the reconstructed Gaussian
channel. Let $X(t)$ and $Y(t)$ be defined through Eqs.\eqref{eq:X-definition} and \eqref{eq:Y-definition}. Moreover, a Gaussian channel is CP if
Eq.\eqref{eq:gaussian-cp} is satisfied.
We therefore introduce the penalty
\begin{equation}
 R_{\mathrm{phys}}(\Theta)=
\sum_{k=1}^{K}\left\|\left[
-\left(Y(t_k)+\frac{i}{2}\Omega-\frac{i}{2}X(t_k)\Omega X^{T}(t_k)\right)\right]_{+}\right\|_{F}^{2}
 \label{eq:physicality-regularizer}
\end{equation}
where $[M]_{+}$ denotes the positive part of the Hermitian matrix $M$. This penalty vanishes when the reconstructed
map is CP at all sampled times. 

For numerical implementation, the time-dependent quantities may be expanded
in a finite set of basis functions
\begin{align}
A(t)&=\sum_{\ell=1}^{N_A}A_{\ell} f_{\ell}(t)
\label{eq:A-basis} \\
D(t)&=\sum_{\ell=1}^{N_D}D_{\ell} g_{\ell}(t)\label{eq:D-basis} \\
\bm u(t)&=\sum_{\ell=1}^{N_u}\bm u_{\ell} h_{\ell}(t)
\label{eq:u-basis-expansion}
\end{align}
where $N_A$, $N_D$, and $N_u$ denote the numbers of basis functions used
for the drift, diffusion, and displacement-drift terms, respectively and $f_{\ell}(t)$, $g_{\ell}(t)$, and $h_{\ell}(t)$ are scalar basis functions and may be splines, Fourier modes, exponentials, or hardware-motivated memory functions. The unknown optimization vector is obtained by collecting the independent
entries of the coefficient matrices and vectors
\begin{equation}
\Theta=\left\{A_{\ell}\right\}_{\ell=1}^{N_A}
\cup
\left\{D_{\ell}\right\}_{\ell=1}^{N_D}
\cup
\left\{\bm u_{\ell}\right\}_{\ell=1}^{N_u}
 \label{eq:parameter-vector}
\end{equation}
together with any additional parameters appearing in the Hamiltonian or in the
noise model. The coefficients $A_\ell$ and $\bm u_\ell$ are real,
and each $D_\ell$ is real symmetric. When fitting an
unrestricted drift matrix $A(t)$, its coherent contribution
is already included; hence $G(t)$ is not fitted separately.
A separate Hamiltonian parameterization requires a fixed
decomposition of the drift to avoid redundancy.

The reconstruction is identifiable only to the extent allowed by the chosen probe states, measurement times, and homodyne directions. In practice, the input Gaussian states and measured quadratures must be sufficiently rich to distinguish the unknown parameters of the selected generator model. For this reason, we use a finite-dimensional, physically motivated parameterization of the time-dependent noise rates instead of attempting to reconstruct arbitrary matrix-valued functions $A(t)$, $D(t)$, and $\bm u(t)$. The reliability of the reconstruction is assessed through the conditioning of the likelihood landscape and through predictive model comparison on test synthetic data. The reconstructed time-local generator then provides the input for the CP-divisibility analysis carried out in the next section.

\section{From Reconstructed Dynamics to Memory Detection}
\label{sec:memory_detection}
\subsection{Diagnosing non-Markovianity from the reconstructed generator}
\label{subsec:diagnosing_nonmarkovianity}
After the tomographic reconstruction, the inferred time-local Gaussian
generator is used to decide whether the observed dynamics is compatible
with a Markovian Gaussian Lindblad description or requires a non-Markovian interpretation. In this work, non-Markovianity is diagnosed
at the generator level through CP-divisibility, rather than through
state-dependent distinguishability measures.

The first step is to verify that the reconstructed finite-time Gaussian
channel is physical. This is done by checking the CP
condition for the total Gaussian map, given in Eq.\eqref{eq:gaussian-cp} at all sampled
times. This global physicality check is necessary because a temporary violation of divisibility is meaningful only if the overall reconstructed
map remains CP. The second step is to test the infinitesimal CP-divisibility condition
using Eq.\eqref{eq:infinitesimal-cp}. If this condition is satisfied for all times, the
reconstructed dynamics is consistent with a CP-divisible Gaussian
Lindblad process. If, however, this condition is violated over a finite time
interval while the total Gaussian channel still satisfies CP condition, then
the dynamics is identified as non-Markovian in the CP-divisibility sense
\cite{Rivas2014,Torre2018}.
Operationally, we compute the eigenvalues of the local CP matrix
$M_{\rm CP}(t)$ defined in Eq.\eqref{eq:infinitesimal-cp}. Negative eigenvalues indicate time intervals in which the intermediate map from $t$ to $t+dt$ is not CP.

To quantify the total strength of the violation,
we introduce a CP-divisibility non-Markovianity score,
inspired by the Gaussian-channel non-Markovianity
analysis of Torre et al. \cite{Torre2015}
\begin{equation}
\mathcal N_{\rm CP}=\int_0^T \sum_{\lambda_\ell(t)<0}
\left|\lambda_\ell(t)\right|\,dt 
\label{eq:NCP_score}
\end{equation}
Here $\lambda_\ell(t)$ are the real eigenvalues of the Hermitian matrix $M_{\mathrm{CP}}(t)$, 
$\ell=1,\ldots,2m$ labels the eigenvalues at each time and $T$ is the final reconstruction
time. At each time, the sum includes only the negative eigenvalues. Therefore, $\mathcal N_{\rm CP}=0$ for CP-divisible Gaussian dynamics, whereas $\mathcal N_{\rm CP}>0$ quantifies an accumulated violation of CP-divisibility over a time set of nonzero duration.

For discretely sampled data, Eq.\eqref{eq:NCP_score} is evaluated
numerically as a sum over the reconstruction times
\begin{equation}
\mathcal N_{\rm CP} \simeq \sum_{k=1}^{K-1} \Delta t_k \sum_{\lambda_\ell(t_k)<0}
\left|\lambda_\ell(t_k)\right|,
\qquad
\Delta t_k=t_{k+1}-t_k 
\end{equation}
where $\Delta t_k$ is the time step associated with the sampled point
$t_k$ and $k$ is the index labeling the discrete reconstruction times. This quantity measures the accumulated negative part of the
spectrum of $M_{\rm CP}(t)$.

In numerical reconstructions, small negative eigenvalues may arise from
finite sampling noise or optimization error. Therefore, a non-Markovian
interpretation should be assigned only when the violation of Eq.\eqref{eq:infinitesimal-cp}, or
equivalently the positive value of $\mathcal N_{\rm CP}$ is larger than the estimated reconstruction uncertainty. In practice, this can be
assessed using bootstrap resampling, confidence intervals, or repeated synthetic data sets. Here, the reconstruction uncertainty refers to the statistical and numerical
uncertainty in the reconstructed eigenvalues of $M_{\rm CP}(t)$, arising from
finite homodyne sampling, regularization and numerical optimization.

\subsection{Candidate Markovian and non-Markovian Gaussian noise models}
\label{sec:model_hierarchy}
The tomography procedure described above reconstructs an effective time-local
Gaussian noise generator from time-resolved homodyne data. However, observing time-dependent noise
parameters does not automatically imply that the dynamics is genuinely non-Markovian. A time-dependent but CP-divisible generator may still describe a memoryless process with externally varying control parameters. Therefore, it is
important to compare a collection of candidate models with increasing physical
complexity.

The purpose of this subsection is to compare candidate Gaussian noise models with different levels of physical complexity and to determine which model gives the best predictive description of the observed or simulated homodyne
quadrature data without overfitting. We consider four models. The first is a stationary Markovian baseline. The second allows time-dependent rates but still enforces CP-divisibility. The third allows Lindblad-like non-Markovian dynamics
through temporary violation of the infinitesimal CP condition. The fourth uses an
enlarged Markovian system with auxiliary memory modes, thereby providing a microscopic interpretation of the observed memory effects.

\subsubsection{Model $M_0$: Stationary CP-divisible Gaussian noise}
The simplest model assumes that the noise is stationary and memoryless. In this
case, the Gaussian generator is time independent:
\begin{equation}
 A(t)=A_0,
\qquad
D(t)=D_0,
\qquad
\bm{u}(t)=\bm{u}_0
\label{eq:M_0-static}
\end{equation}
The infinitesimal CP condition must hold $(M_{\mathrm{CP}} \geq 0)$.
This model corresponds to ordinary Markovian Lindblad noise with constant
photon-loss, heating, and diffusion rates. It is the appropriate description when the environment has no memory and the hardware parameters are stable over the experimental time window.
This model should be accepted if it already explains the data within statistical
uncertainty. If it fails, then either the noise is time-dependent, correlated or non-Markovian.

\subsubsection{Model $M_1$: Time-dependent CP-divisible Gaussian noise}
The second model allows the generator to vary in time
\begin{equation}
\begin{aligned}
A(t),\; D(t),\;\bm{u}(t)
\quad
\mathrm{time\ dependent}\\
M_{\mathrm{CP}}(t)\ge 0
\quad \text{for all } t\in[0,T]
\label{eq:M_1-time-dependent}
\end{aligned}
\end{equation}
but still requires the infinitesimal CP condition at all times.
This model describes noise whose strength changes in time but remains CP-divisible. Physically, it can represent slow calibration drift, controlled
changes in optical loss, or externally modulated Gaussian noise. Although the
coefficients depend on time, the dynamics can still be interpreted as Markovian
in the CP-divisibility sense because every intermediate map is CP. This model is useful for separating ordinary time-dependent hardware drift from genuine memory effects. If $M_1$ fits the data well, there is no need to invoke
non-Markovian information backflow.

\subsubsection{Model $M_2$: Time-local non-CP-divisible Gaussian noise}
The third model allows the reconstructed generator to violate CP-divisibility:
\begin{equation}
\begin{aligned}
A(t),\quad &D(t),\quad \bm{u}(t)\quad\mathrm{time\ dependent}, \\
&M_{\mathrm{CP}}(t)\not\geq 0\quad\mathrm{for\ some}\;t \in[0,T]
\label{eq:M_2-noncp}
\end{aligned}
\end{equation}
Equivalently, in the operator form of the generator, the Kossakowski-like
matrix may have temporarily negative eigenvalues, as in Eq.\eqref{eq:non-Markovianity}. This model is called Lindblad-like because
it retains the operator structure of a Lindblad generator, but relaxes the
requirement that all instantaneous rates be positive. Physically, it can describe optical modes coupled to a
structured reservoir or a delayed feedback path, where
temporary storage and return of excitations can produce
non-CP-divisible reduced dynamics. The total map from the
initial time to the final time must still remain physical, satisfying
Eq.\eqref{eq:gaussian-cp}. Therefore, $M_2$ describes a situation in which
the full evolution is CP, while the intermediate evolution is not CP-divisible. In the present framework, this combination of a physical
finite-time Gaussian channel with a non-CP-divisible time-local generator is
the main signature of non-Markovian memory. This model should be preferred
over $M_1$ only if the data require negative instantaneous rates or violation
of Eq.\eqref{eq:infinitesimal-cp} to obtain a significantly better fit.

Note that Eq.\eqref{eq:M_2-noncp} defines the non-CP-divisible regime.
In the benchmarks, the label $M_2$ denotes the broader
finite-time CP fitting family in which local CP violations
are allowed but not imposed; a fitted member belongs
to the regime of Eq.\eqref{eq:M_2-noncp} only when such a violation
occurs.

\subsubsection{Model $M_3$: Pseudomode-embedded Markovian model}
The fourth model gives a microscopic explanation for non-Markovianity by
embedding the photonic system into a larger Markovian system. The physical
photonic modes $a_j$ are coupled to auxiliary memory modes $b_\ell$ through
Eq.\eqref{eq:embedded_hamiltonian}. The enlarged density operator
$\varrho(t)$ evolves according to the Markovian Lindblad equation in
Eq.\eqref{eq:embedded_lindblad}, while the observed photonic state is obtained by tracing out the memory modes, $\rho(t)=\mathrm{Tr}_{b}[\varrho(t)]$. Although the enlarged dynamics is Markovian, the reduced photonic dynamics may be non-Markovian. 
This model is physically motivated when the photonic
circuit couples to auxiliary degrees of freedom, such as
ring resonators, delay modes, or spectral side modes,
that retain information about its earlier evolution.
Their subsequent influence on the circuit can generate
memory effects, although their presence alone does not
guarantee non-CP-divisible dynamics.
Compared with the
phenomenological time-local model $M_2$, the pseudomode model $M_3$ is more
structured because it attributes the observed memory effects to explicit auxiliary modes.

\subsection{Statistical model selection}
After fitting the candidate models, we compare them using predictive performance and information criteria. This step is necessary because a more flexible time-dependent generator can always improve the training likelihood, even when the improvement is not statistically meaningful. Model selection therefore determines whether the data justify moving from a stationary Markovian description to a time-dependent CP-divisible model, or further to a non-CP-divisible time-local model.

Let $\Theta_M$ be the fitted parameters of model $M$, $\nu_M$ denotes the number of fitted parameters in each model, $N$ is the
number of homodyne samples, and $\mathcal J_M(\Theta_M)$ denotes the
negative log-likelihood of the data for model $M$ in Eq.\eqref{eq:negative-log-likelihood}, evaluated at the fitted parameters and excluding the regularization penalties. We use the Akaike
information criterion (AIC) and Bayesian information criterion (BIC) as standard model-selection criteria that penalize goodness of fit by the
number of fitted parameters \cite{Akaike1974,Schwarz1978,Dziak2020,Guta2012}
\begin{equation}
{\rm AIC}_M= 2\nu_M+2\mathcal J_M(\Theta_M)
\label{eq:AIC}
\end{equation}
\begin{equation}
{\rm BIC}_M =\nu_M\log N+2\mathcal J_M(\Theta_M)
\label{eq:BIC}
\end{equation}
Lower values of $\mathrm{AIC}_M$ and $\mathrm{BIC}_M$
indicate a better balance between how well the model fits the data and the size of its parameterization.
The regularizers influence the fitted parameters
$\Theta_M$ and therefore indirectly affect the value
of $\mathcal{J}_M({\Theta}_M)$. However, their numerical values are not added directly to the information criteria: $\mathrm{AIC}_M$ and $\mathrm{BIC}_M$ use the data negative log-likelihood evaluated at the fitted parameters, together with their respective $\nu_M$-dependent terms.

We also evaluate the predictive negative log-likelihood on held-out test data
\begin{equation}
\begin{aligned}
\epsilon_M^{\mathrm{test}}&=\frac{1}{N_{\rm test}}\mathcal J_M^{\rm test}(\Theta_M)\\
&= \frac{1}{N_{\rm test}}
\sum_{n\in{\rm test}}
-\log p(x_n|j_n, \theta_n,t_n,\alpha_n;\Theta_M)
\label{eq:test-nll}
\end{aligned}
\end{equation}
Here $p$ is the model-predicted probability density for one homodyne outcome $x_n$ and $\epsilon_M^{\mathrm{test}}$ is the average negative
log-likelihood per test sample. A smaller value means that the fitted model assigns, on average, higher probability to homodyne outcomes that were not used during parameter estimation, and therefore has better predictive performance.
This comparison helps determine whether an improvement on the training data reflects genuine predictive structure or merely overfitting.

Considering these criteria, we compare the candidate Gaussian noise models according to both their statistical support and their physical interpretation. 

\subsection{Algorithm for non-Markovian Gaussian noise reconstruction}
\label{subsec:algorithm}
The reconstruction framework introduced above defines the inverse problem at
the level of homodyne probability distributions and Gaussian moment dynamics. To apply this framework to simulated or experimental data, we summarize the procedure as an explicit algorithm. The purpose of the algorithm is to convert
time-resolved quadrature measurements into an estimate of the time-local
Gaussian generator, and then to test whether the reconstructed generator is
compatible with CP-divisible Markovian dynamics or shows signatures of non-CP-divisible non-Markovian behavior.

Algorithm \ref{alg:NM_Gaussian_tomography} summarizes the operational implementation of the proposed tomographic reconstruction workflow. The
algorithm has three conceptual stages. First, one specifies the probe states, measurement times, homodyne angles, and a parameterized model for the unknown Gaussian generator. Second, for each trial parameter set, the Gaussian moment equations are solved and the likelihood of the observed data is evaluated.
Third, the optimized generator is checked for physical consistency of the finite-time Gaussian channel and then used to test CP-divisibility through
$M_{\rm CP}(t)$ and the corresponding non-Markovianity score $\mathcal N_{\rm CP}$. 

\begin{algorithm*}[t]
\centering
\refstepcounter{algorithm}
\label{alg:NM_Gaussian_tomography}
\begin{tcolorbox}[
    enhanced,
    width=\textwidth,
    colback=white,
    colframe=black!60,
    colbacktitle=black!35,
    coltitle=black,
    fonttitle=\bfseries,
    halign title=center,
    title={Algorithm~\thealgorithm: Tomographic reconstruction and CP-divisibility diagnosis of Gaussian noise dynamics},
    boxrule=0.8pt,
    arc=2mm,
    outer arc=2mm,
    left=8pt,
    right=8pt,
    top=7pt,
    bottom=7pt,
    titlerule=0pt,
    toptitle=5pt,
    bottomtitle=5pt
]

\begin{algorithmic}[1]

\STATE Choose a set of Gaussian input probe states
$\{\rho_{\alpha}(0)\}$.

\STATE Choose measurement times $\{t_k\}$
and homodyne settings $\{X_{j,\theta}\}$.

\STATE Collect the time-resolved homodyne data $\mathcal{D}=\{x_n,j_n,\theta_n,t_n,\alpha_n\}_{n=1}^{N}.
$

\STATE Split $\mathcal{D}$ into training and test sets,
$\mathcal{D}_{\mathrm{tr}}$ and
$\mathcal{D}_{\mathrm{test}}$.

\FOR{each candidate model
$M\in\{M_0,M_1,M_2,M_3\}$}

    \STATE Parameterize the candidate dynamics through the model parameters
    $\Theta_M$, either directly through $A_M(t)$, $D_M(t)$ and $\bm{u}_M(t)$,
    or through an equivalent reduced Gaussian channel or pseudomode embedding.

    \STATE For each numerical evaluation of $\Theta_M$,
    propagate $\mathbf{d}_{\alpha}(t)$ and
    $V_{\alpha}(t)$ using Eqs.\eqref{eq:first-moment} and
    \eqref{eq:second-moment}, or an analytically equivalent
    Gaussian-channel propagation.

    \STATE Compute the predicted homodyne means and variances using
    Eqs.\eqref{eq:mean} and \eqref{eq:variance}.

    \STATE Evaluate the training negative log-likelihood in
    Eq.\eqref{eq:negative-log-likelihood}.

    \STATE Minimize the training objective, using
    Eq.\eqref{eq:regularized-estimator} when regularization is required,
    and impose the physical constraints of the selected model.

    \STATE Verify CP of the reconstructed finite-time channel using
    Eq.\eqref{eq:gaussian-cp}.

    \STATE Compute $M_{\rm CP}(t)$ using
    Eq.\eqref{eq:infinitesimal-cp} and evaluate the CP-divisibility witness
    $\lambda_{\min}[M_{\rm CP}(t)]$.

    \STATE Compute $\mathrm{AIC}_M$, $\mathrm{BIC}_M$ and
    $\epsilon_M^{\mathrm{test}}$ using Eqs.\eqref{eq:AIC},
    \eqref{eq:BIC} and \eqref{eq:test-nll}, respectively.

\ENDFOR

\STATE Compare the candidate models using
$\mathrm{AIC}_M$, $\mathrm{BIC}_M$ and
$\epsilon_M^{\mathrm{test}}$, and select the best-supported physically valid model.

\STATE Estimate reconstruction uncertainty by generating
parametric bootstrap samples from the fitted selected
model, refitting its parameters and constructing
confidence intervals for $\lambda_{\min}[M_{\rm CP}(t)]$.

\STATE Report the selected model.

\end{algorithmic}
\end{tcolorbox}
\end{algorithm*}

\section{Numerical validation}
\label{sec:numerical_validation}
We validate the framework using synthetic homodyne data
generated from known Gaussian dynamics. The benchmarks
represent distinct noise mechanisms rather than a specific
device, allowing direct comparison with the reference
dynamics and testing the distinction between time-dependent
CP-divisible noise and non-CP-divisible memory. In these benchmarks, source attribution is conditional
on the specified model families and environmental
assumptions; it does not establish a unique microscopic
identification of the noise mechanism. Candidate
models are fitted to training data and compared using
information criteria and predictive performance on
independent test data. Numerical details are provided
in Appendix \ref{app:numerical_details}.
All simulations, data generation,
parameter fitting, and bootstrap analyses were implemented
in Python using NumPy and SciPy.

\subsection{Benchmark 1: single-mode squeezed state under memory-dependent loss}
\label{sec:benchmark1_revised}
The first benchmark considers a single-mode squeezed vacuum state with zero displacement and covariance matrix \cite{Weedbrook2012}
\begin{equation}
V(0)=\frac{1}{2}
\begin{pmatrix}
e^{-2r} & 0\\
0 & e^{2r}
\end{pmatrix}
\label{eq:b1_initial_cov}
\end{equation}
The squeezing parameter $r=0.80$ is chosen as a benchmark input parameter rather
than fitted from the data and corresponds to approximately
$10\log_{10}(e^{2r})\simeq 6.95\,{\rm dB}$ of quadrature squeezing, giving a
moderately squeezed input state whose variance is sensitive to memory-dependent loss.  
\subsubsection{Candidate models}
The reference loss rate is chosen as a positive background
loss rate $\kappa_0$ plus a damped oscillatory correction
with amplitude $\alpha_\kappa$, memory decay time $\tau_c$,
and oscillation frequency $\omega_c$.
A physical motivation is an optical mode subject to
background loss and coupled to a damped auxiliary resonator:
coherent excitation exchange can produce oscillations in
the effective loss rate, while resonator damping suppresses
these oscillations over time. The coupling between the modes
need not vary in time; a time-dependent reduced loss rate
can arise even for constant coupling. We represent this behavior through the following
phenomenological effective loss rate, without assuming
an exact correspondence to a specific mode-resonator
system \cite{Breuer2007}
\begin{equation}
\kappa_{\rm ref}(t) =
\kappa_0+\alpha_{\kappa}e^{-t/\tau_c}\cos(\omega_c t)
\label{eq:b1_kappa_ref}
\end{equation}
with
\begin{equation}
\kappa_0=0.08,
\qquad
\alpha_{\kappa}=0.18,
\qquad
\tau_c=3.0,
\qquad
\omega_c=2.2
\label{eq:b1_ref_params}
\end{equation}
The parameters are chosen such that $\kappa_{\rm ref}(t)$
becomes temporarily negative between its two zero crossings,
producing a violation of CP-divisibility, while the accumulated
loss (defined below) remains non-negative so that the
finite-time Gaussian channel remains CP and physical.
This non-negativity is verified over the complete observation
interval by checking its endpoints and all stationary points
of the accumulated loss.

The time-dependent loss rate produces the accumulated loss
\begin{equation}
\Gamma_{\rm ref}(t)=
\int_0^t \kappa_{\rm ref}(s)\,ds
\label{eq:b1_gamma_ref}
\end{equation}
and the transmissivity of the finite-time pure loss channel is therefore
\begin{equation}
\eta_{\rm ref}(t)=e^{-\Gamma_{\rm ref}(t)},
\qquad
0\leq \eta_{\rm ref}(t)\leq 1
\label{eq:b1_eta_ref}
\end{equation}

A pure-loss Gaussian channel is equivalent to mixing the input mode with a vacuum environmental mode at a beam splitter of transmissivity $\eta(t)$. Using $V_{\rm vac}=I_2/2$, the covariance matrix therefore
evolves as \cite{Weedbrook2012}
\begin{equation}
V(t)=\eta(t)V(0)+
\left[1-\eta(t)\right]\frac{I_2}{2}
\label{eq:covariance_evolution}
\end{equation}
To make the comparison between the time-dependent CP-divisible and non-CP-divisible descriptions as controlled as possible, models $M_1$ and $M_2$ are assigned the same parameter temporal family
\begin{equation}
\kappa_M(t)=\kappa_{0,M}+
\alpha_M e^{-t/\tau_M}\cos(\omega_M t)
\label{eq:b1_common_family}
\end{equation}
Thus, $M_1$ and $M_2$ differ only through their physical constraints rather than through different fitting functions.
The time-dependent CP-divisible model $M_1$ uses Eq.\eqref{eq:b1_common_family} subject to
\begin{equation}
\kappa_{M_1}(t)\geq 0
\quad \text{for all }t\in[0,T]
\label{eq:b1_M1}
\end{equation}
The time-local model $M_2$ uses the same
Eq.\eqref{eq:b1_common_family}, but allows the
instantaneous loss rate to take either positive or
negative values while enforcing finite-time physicality.
\begin{equation}
\Gamma_{M_2}(t)=\int_0^t\kappa_{M_2}(s)\,ds
\geq 0
\quad \text{for all }t\in[0,T]
\label{eq:b1_M2}
\end{equation}
Models $M_1$ and $M_2$ share the same temporal ansatz and parameter count, but allowing negative rates enlarges the admissible parameter region of $M_2$. Consequently, an improved likelihood indicates better relative model support but does not independently establish the statistical necessity of negative rates. Since $M_1$ is nested within $M_2$ with a boundary constraint, the information criteria are interpreted as descriptive model comparisons rather than formal significance tests.

The stationary model $M_0$ assumes a constant Markovian non-negative loss rate
\begin{equation}
\kappa_{M_0}(t)=\kappa_{M_0}
\label{eq:b1_M0}
\end{equation}

Model $M_3$ is conceptually different from $M_0$-$M_2$.
It does not postulate a phenomenological time-dependent loss rate for the observed mode.  Instead, the observed photonic mode $a$ is coupled to an
auxiliary bosonic memory mode $b$ and the joint system is assumed to evolve according to a Markovian Lindblad equation.  This is the standard
pseudomode mechanism by which non-Markovian reduced dynamics can arise
from Markovian dynamics on an enlarged Hilbert space
\cite{Garraway1997,Dalton2001,Pleasance2020}.

For Benchmark 1, we use one resonant auxiliary mode, initially in the
vacuum state and based on Eq.\eqref{eq:embedded_lindblad} we write
\begin{equation}
\frac{d\varrho(t)}{dt}=-i\left[g
\left(a^\dagger b+a b^\dagger\right),
\varrho(t)
\right]+\gamma_m\,
\mathcal{D}[b]\varrho(t)
\label{eq:b1_M3_master}
\end{equation}
where $\varrho(t)$ is the joint density operator, $\gamma_m\geq0$ is the damping rate of the
auxiliary memory mode, and $g\geq0$ is the real coherent
system-memory coupling.  The decay rates in
Eq.\eqref{eq:b1_M3_master} are non-negative, so the joint
two-mode dynamics is Markovian and CP.

The corresponding Heisenberg-Langevin equations are
\cite{Gardiner1985}
\begin{align}
\dot a(t)
&=-i g b(t)
\label{eq:b1_M3_a}
\\
\dot b(t)
&=-\frac{\gamma_m}{2}b(t)-i g a(t)
+\sqrt{\gamma_m}\,
b_{\rm in}(t)
\label{eq:b1_M3_b}
\end{align}
where $b_{{\rm in}}(t)$ is the input-field operator of the independent Markovian vacuum reservoirs coupled to the pseudomodes.
Because the input reservoirs are in vacuum, their mean values vanish.
The first moments therefore obey
\begin{equation}
\frac{d}{dt}
\begin{pmatrix}
\langle a(t)\rangle\\[1mm]
\langle b(t)\rangle
\end{pmatrix}
=
\mathsf{P}
\begin{pmatrix}
\langle a(t)\rangle\\[1mm]
\langle b(t)\rangle
\end{pmatrix}
\label{eq:matrix}
\end{equation}
where
\begin{equation}
\mathsf{P}=
\begin{pmatrix}
0 & -i g\\
-i g & -\gamma_m/2
\end{pmatrix}
\label{eq:b1_M3_first_moments}
\end{equation}
Since the auxiliary mode is initially in vacuum,
$\langle b(0)\rangle=0$, the solution of Eq.\eqref{eq:matrix} gives
\begin{equation}
\langle a(t)\rangle=
\chi_3(t)\langle a(0)\rangle,
\qquad
\chi_3(t)
=\left[e^{\mathsf{P}t}\right]_{11}
\label{eq:b1_M3_transfer_definition}
\end{equation}
Here, the subscript $11$ denotes the first-row, first-column element of the matrix exponential, corresponding to the transfer amplitude from the initial system mode $a(0)$ back to the observed mode $a(t)$.
For the resonant two-mode model in
Eq.\eqref{eq:b1_M3_first_moments}, the matrix exponential can be
evaluated analytically. Using Eqs.\eqref{eq:b1_M3_a} and \eqref{eq:b1_M3_b}, the auxiliary-mode amplitude
$\langle b(t)\rangle$ can be eliminated, giving
\begin{equation}
\langle \ddot{a}(t)\rangle
+\frac{\gamma_m}{2}{\langle \dot{a}(t)\rangle}
+g^2\langle a(t)\rangle=0 
\end{equation}
The two characteristic exponents are
$-\gamma_m/4\pm\Delta$, where we define the pseudomode spectral
discriminant
\begin{equation}
\Delta=\sqrt{\frac{\gamma_m^2}{16}
-g^{\,2}}
\label{eq:b1_M3_delta}
\end{equation}
which determines whether the embedded dynamics is overdamped or oscillatory.
Using $\chi_3(0)=1$ and $\dot{\chi}_3(0)=0$, which follow from
$\langle b(0)\rangle=0$, the general solution is
\begin{equation}
\chi_3(t)= e^{-\gamma_m t/4}
\left[
\cosh\left(\Delta t\right)
+\frac{\gamma_m}{4\Delta}
\sinh\left(\Delta t\right)
\right]
\label{eq:b1_M3_transfer}
\end{equation}
At critical damping ($\Delta=0$), this expression is
understood through its continuous limit,
$\chi_3(t)=e^{-\gamma_m t/4}(1+\gamma_m t/4)$.
For
$g>\frac{\left|\gamma_m\right|
}{4}$, $\Delta$ is imaginary.  Writing
\begin{equation}
\omega_{\rm{osc}}=\sqrt{g^{\,2}
-\frac{\gamma_m^2
}{16}}
\label{eq:b1_M3_Omega}
\end{equation}
as the oscillation frequency generated by the coupling between the system mode $a$ and the pseudomode. So, 
Eq.\eqref{eq:b1_M3_transfer} becomes
\begin{equation}
\chi_3(t)=
e^{-\gamma_m t/4}\left[
\cos \left(\omega_{\rm{osc}} t\right)
+\frac{\gamma_m}{4\omega_{\rm{osc}}}
\sin \left(\omega_{\rm{osc}} t\right)\right]
\label{eq:b1_M3_transfer_oscillatory}
\end{equation}
Thus, the damped oscillations appearing in the reduced dynamics of $M_3$ arise from coherent excitation exchange with the auxiliary
memory mode. This mode and its Markovian reservoir are
initially in vacuum. The coupling between the
system and memory modes is passive; since $\chi_3(t)$ is the field-amplitude transfer coefficient in Eq.\eqref{eq:b1_M3_transfer_definition}, while
transmissivity refers to the corresponding intensity attenuation, the
reduced quantum-limited attenuation channel has transmissivity
\begin{equation}
\eta_3(t)=\left|\chi_3(t)\right|^2 
\label{eq:b1_M3_eta}
\end{equation}
Consequently, the covariance matrix of the observed mode evolves as Eq.\eqref{eq:covariance_evolution} with $\eta_3(t)$. This is the quantity used to predict the homodyne variances and hence
to evaluate the likelihood of model $M_3$.

To compare the embedded model with the reduced loss-rate descriptions $M_0$-$M_2$, we define the effective reduced loss rate based on Eq.\eqref{eq:b1_eta_ref}
\begin{equation}
\kappa_3^{\rm eff}(t)
=-\frac{d}{dt}\ln\eta_3(t)
\label{eq:b1_M3_effective_rate_definition}
\end{equation}
which, using Eq.\eqref{eq:b1_M3_eta} gives
\begin{equation}
\kappa_3^{\rm eff}(t)=-2\,\mathrm{Re}
\left[\frac{\dot\chi_3(t)}{\chi_3(t)}
\right]
\label{eq:b1_M3_effective_rate}
\end{equation}
Eq.\eqref{eq:b1_M3_effective_rate} is defined on intervals for
which $\chi_3(t)\neq0$.  A temporary increase of the reduced
transmissivity $\dot\eta_3(t)>0$ is equivalent to $\kappa_3^{\rm eff}(t)<0$,
and therefore signals temporary information or excitation return from
the memory mode to the observed mode.  The reduced evolution can thus
become non-CP-divisible even though the joint evolution in
Eq.\eqref{eq:b1_M3_master} remains a CP Markovian Lindblad dynamics.
Importantly, $\kappa_3^{\rm eff}(t)$ is a derived quantity and is not assigned
the damped-cosine form of Eq.\eqref{eq:b1_common_family}. Model $M_3$
is therefore statistically compared with $M_0$--$M_2$ at the level of
the same measured homodyne data, while representing a genuinely
different physical hypothesis for the origin of the observed memory.

\subsubsection{Fit results and uncertainty assessment}
The resulting fits are compared using the information criteria and held-out predictive score introduced above, together with a bootstrap assessment of the reconstructed CP-divisibility violation.
Each candidate model $M_0$-$M_3$ was fitted to the training data by minimizing the training
negative log-likelihood in Eq.\eqref{eq:negative-log-likelihood}, using numerical optimization routines implemented in
Python with NumPy/SciPy. 
For these low-dimensional benchmark models, no regularization is required and the parameters are obtained by minimizing the training negative log-likelihood subject to the model-specific physicality constraints. So, after optimization the unregularized training negative log-likelihood was used to compute
the AIC and BIC values in Eqs.\eqref{eq:AIC} and \eqref{eq:BIC}, the corresponding
relative scores $\Delta{\rm AIC}$ and $\Delta{\rm BIC}$ were obtained
from Eq.\eqref{eq:delta_AIC}, and the predictive negative log-likelihood was
evaluated independently on the held-out test data using Eq.\eqref{eq:test-nll}.
\begin{table}[t]
\centering
\caption{Model comparison for benchmark 1. Relative AIC
and BIC are reported with respect to the best model; $\epsilon_M^{(\mathrm{test})}$ is reported per test sample and $\nu_M$ denotes the number of independent real
parameters in model $M$.}
\label{tab:validation_single_mode_results}
\small
\setlength{\tabcolsep}{11pt}      
\renewcommand{\arraystretch}{1.9} 
\begin{tabular}{|c c c c c|}
\hline
Model & $\nu_M$ & $\Delta\mathrm{AIC}$ & $\Delta\mathrm{BIC}$ & $\epsilon_M^{(\mathrm{test})}$ \\
\hline
$M_0$ & 1 & 255.3 & 221.9 & 1.31612 \\
$M_1$ & 4 & 14.0  & 14.0  & 1.31586 \\
$M_2$ & 4 & 0.0   & 0.0   & 1.31584 \\
$M_3$ & 2 & 981.0   & 958.7   & 1.31682 \\
\hline
\end{tabular}
\end{table}

The results in Table \ref{tab:validation_single_mode_results} were obtained from the numerical reconstruction pipeline described above in which $\nu_M$ counts the fitted real parameters: $\kappa_{M_0}$ for $M_0$; $(\kappa_{0,M},\alpha_M,\tau_M,\omega_M)$ for $M_1$ and $M_2$; and $(g,\gamma_m)$ for the resonant pseudomode model $M_3$.
As shown, the time-local model $M_2$ is selected
by both information criteria, with $\Delta{\rm AIC}=\Delta{\rm BIC}=0$,
and also gives the lowest test score. Importantly, $M_1$ and $M_2$ use the same four-parameter damped-oscillatory family. The preference for $M_2$ is consistent with its ability to capture the temporary negative-rate interval excluded by $M_1$, although this preference alone does not establish the statistical necessity of non-CP-divisible dynamics. The stationary model $M_0$ is strongly
disfavored because it cannot reproduce the temporal memory structure. The pseudomode model $M_3$ is also strongly disfavored. Its fitted
parameters place the pseudomode dynamics in the overdamped regime, so
the effective loss builds up monotonically through coupling to the
damped memory mode and does not reproduce the oscillatory negative-rate
structure of the reference process. This is consistent with the fact that the synthetic data in this benchmark are generated from the signed time-local family rather than from a pseudomode-embedded dynamics.
\begin{figure}
\centering
\includegraphics[width=0.93\linewidth]{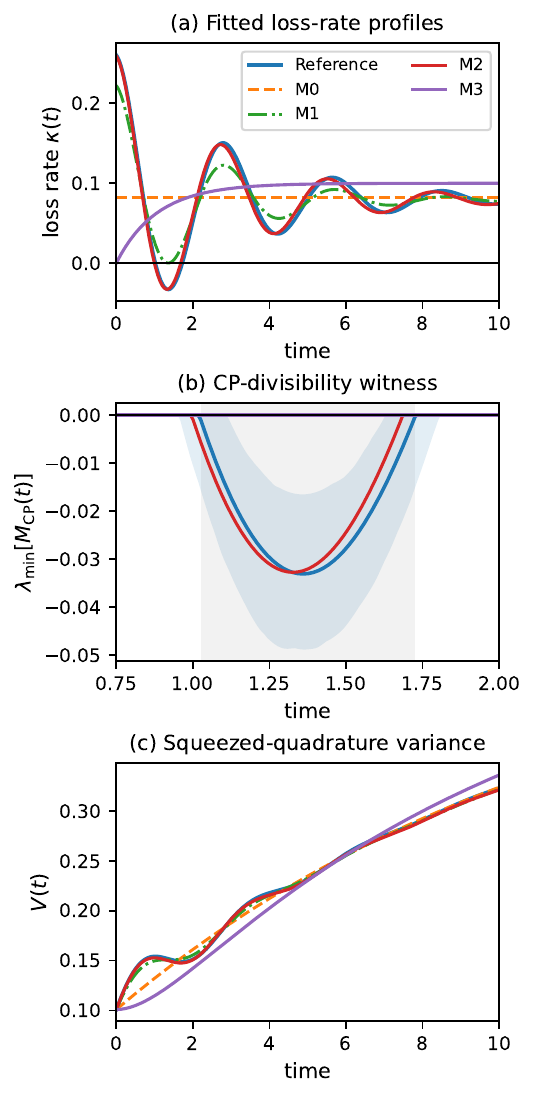}
\caption{Benchmark 1: single-mode squeezed state under memory-dependent loss.  (a) Reference loss rate in Eq.\eqref{eq:b1_kappa_ref}, fitted $M_0$-$M_2$ rates, and the effective reduced loss rate derived from the fitted pseudomode model $M_3$.  Models $M_1$ and $M_2$ use the same damped-cosine family, with $M_1$ constrained to remain CP-divisible.  (b) Smallest eigenvalue of the infinitesimal Gaussian CP matrix.  Negative values identify the interval of non-CP-divisible reduced dynamics; the light band shows the $95\%$ pointwise bootstrap uncertainty for $M_2$.  (c) Squeezed-quadrature variance $V_q(t)$ obtained from the corresponding finite-time transmissivities. The
memory-dependent dynamics produces a small transient recovery of the
variance during the interval of negative effective loss.}
\label{fig:benchmark1}
\end{figure}
To assess the robustness of the reconstructed negative-rate
interval against finite-sampling uncertainty, we perform
a parametric bootstrap with $N_{\rm boot}=300$.
Each synthetic training data set is generated from the
fitted $M_2$ channel, using the same probe states, measurement
times, homodyne settings, and number of shots as in the
original data. Model $M_2$ is independently refitted to each
bootstrap realization, and the resulting distribution is
used to construct confidence intervals for the
reconstructed negative-rate interval. The $95\%$ bootstrap interval for the minimum reconstructed rate is
\begin{equation}
\min_t\kappa_{M_2}(t)
\in [-0.0494,-0.0165]
\end{equation}
which lies entirely below zero. Moreover, the pointwise $97.5\%$ upper bootstrap bound remains negative approximately over
\begin{equation}
t\in[1.12,1.63]
\end{equation}
These bootstrap results support a robust negative-rate interval within the fitted $M_2$ model, although they do not constitute a calibrated hypothesis test against the CP-divisible model $M_1$.

Figure \ref{fig:benchmark1} illustrates the memory-dependent loss dynamics at three different physical levels. 
Panel (a) compares the reference loss rate with the fitted models. The stationary Markovian model $M_0$ captures only the average attenuation, while the CP-divisible model $M_1$ follows part of the temporal variation but cannot enter the negative-rate region because of its positivity constraint. In contrast, the signed time-local model $M_2$ closely reconstructs the damped oscillatory reference profile, including the interval of negative loss rate. The pseudomode model $M_3$ is obtained from an explicit enlarged Markovian dynamics and, for this benchmark, does not reproduce the full oscillatory structure as accurately as $M_2$, which is expected because the synthetic data are generated directly from the time-local reference model. The monotonic $M_3$ profile is physically consistent with the fitted overdamped pseudomode dynamics: the observed mode has no direct loss channel, and its effective attenuation develops only through coherent transfer to the damped pseudomode.

Panel (b) shows the corresponding local CP-divisibility witness $\lambda_{\min}[M_{\rm CP}(t)]$. The witness becomes negative for the reference dynamics and for $M_2$, identifying the interval in which the infinitesimal reduced evolution is not CP-divisible, while $M_0$, $M_1$, and $M_3$ remain non-negative. 
The vertical shaded region marks the time interval in which the reference loss rate is negative, whereas the shaded band around the $M_2$ witness represents its bootstrap uncertainty.
The persistence of the negative region beyond this uncertainty supports a statistically resolved violation of CP-divisibility.

Finally, panel (c) shows the variance $V_q(t)$ of the initially squeezed
$q$ quadrature, with $V_q(0)=e^{-2r}/2$; for notational simplicity, this
quantity is denoted by $V(t)$ in the figure. For a pure-loss channel,
$V_q(t)=\eta(t)V_q(0)+[1-\eta(t)]/2$. Since $V_q(0)<1/2$, ordinary
attenuation decreases $\eta(t)$ and therefore increases $V_q(t)$ toward
the vacuum value $1/2$, corresponding to a gradual degradation of the
initial squeezing.
The reference dynamics and $M_2$ exhibit a small temporary decrease of
$V_q(t)$ during the negative-rate interval. This corresponds to an
increase of the transmissivity and hence to a partial recovery of the
initial squeezing. In contrast, $M_0$ gives a smooth monotonic degradation
determined by its constant effective loss rate, while $M_1$ remains
monotonic because its instantaneous loss rate is constrained to be non-negative. Model $M_3$ also shows no corresponding recovery in this
benchmark. Its fitted overdamped pseudomode dynamics produces a monotonic effective attenuation, so the quadrature variance increases smoothly rather than reproducing the transient recovery present in the
reference dynamics and $M_2$. 

\subsection{Benchmark 2: nonstationary thermal excitation}
Benchmark 1 tests whether the reconstruction can resolve a genuinely non-CP-divisible loss process.  Here we consider the complementary
specificity test: the noise is clearly nonstationary, but the local
Gaussian generator remains CP-divisible throughout the observation
window.  This benchmark therefore tests whether the reconstruction distinguishes ordinary time-dependent thermal excitation from a genuine local breakdown of CP-divisibility.
\subsubsection{Candidate models}
The reference channel is the thermal-loss process introduced
in Eq.\eqref{eq:thermal_master_revised}, with fixed damping
rate $\kappa_{\rm th}=0.20$ and time-dependent bath occupation.
A physical example is a microwave resonator coupled to a
broadband thermal reservoir whose noise level is externally
varied while the coupling strength remains constant.
This changes the reservoir's effective thermal occupation
and hence the noise injected into the resonator, without
changing its damping rate. Assuming that the reservoir
correlation time remains short compared with the relevant
evolution timescales, we model its occupation as
\cite{Breuer2007}
\begin{equation}
\begin{aligned}
&\bar n_{\rm ref}(t)=\bar n_{0,\rm ref}
+\alpha_{\mathrm{ref}} e^{-t/\tau_{\rm ref}}
\cos(\omega_{\rm ref}t)\\
&\bar n_{\rm ref}(t)
=0.30+0.25e^{-t/2.8}\cos(1.8t)
\label{eq:b2_thermal_reference}
\end{aligned}
\end{equation}
No additional dynamical relation is required here: the covariance is propagated directly with Eq.\eqref{eq:second-moment} using the drift and diffusion terms
already given in Eqs.\eqref{photon-loss} and \eqref{eq:D_total_thermal_revised}.  The profile in Eq.(\ref{eq:b2_thermal_reference}) remains positive over the entire
interval; numerically
\begin{equation}
\min_{t\in[0,T]}\bar n_{\rm ref}(t)
\simeq 0.1633
\label{eq:b2_thermal_min}
\end{equation}
For this thermal-loss channel, the two nonzero infinitesimal Kossakowski rates are $\kappa_{\rm th}[\bar n(t)+1]$ and
$\kappa_{\rm th}\bar n(t)$.  Hence the smallest eigenvalue of the
local Gaussian CP matrix is
\begin{equation}
\lambda_{\min}[M_{\rm CP}(t)]=
\kappa_{\rm th}\bar n(t)
\label{eq:b2_thermal_witness}
\end{equation}
which is strictly positive for the reference process.  Time dependence
of the bath occupation therefore does not by itself imply non-Markovianity in the CP-divisibility sense \cite{Breuer2007,Rivas2014}.

The candidate models follow the discussed hierarchy. Model $M_0$ uses a constant thermal occupation and therefore contains one fitted
parameter.  Model $M_1$ preserves the thermal-loss structure of
Eq.\eqref{eq:thermal_master_revised} and uses
\begin{equation}
\bar n_{M_1}(t)=\bar n_{0,M_1}+\alpha_{M_1}e^{-t/\tau_{M_1}}\cos(\omega_{M_1}t),
\quad
\bar n_{M_1}(t)\ge0
\label{eq:b2_M1}
\end{equation}
with four fitted parameters $(\bar n_{0,M_1},
\alpha_{M_1},
\tau_{M_1},
\omega_{M_1})$.  Thus $M_1$ allows nonstationary thermal
noise while enforcing local CP-divisibility. In model $M_1$, the decay and excitation rates are not fitted independently. Because $M_1$ assumes a thermal-loss channel, both are
determined by the same reconstructed bath occupation $\bar n_{M_1}(t)$ with the damping rate $\kappa_{\rm th}$ fixed.

Model $M_2$ is the more general time-local, phase-insensitive Gaussian description. In contrast to $M_1$, it does not impose the
thermal relation between the loss and excitation processes. We denote
by $\kappa_{\downarrow}(t)$ the decay rate multiplying $\mathcal{D}[a]$, which describes photon loss from the system, and by
$\kappa_{\uparrow}(t)$ the excitation rate multiplying
$\mathcal{D}[a^\dagger]$, which describes thermal excitation of the
system by the environment. For an ordinary thermal bath, these rates
are not independent but satisfy
\begin{equation}
\label{eq:thermal_bath}
\kappa_{\downarrow}(t) =
\kappa_{\mathrm{th}}\bigl[\bar n(t)+1\bigr],
\qquad
\kappa_{\uparrow}(t)
=
\kappa_{\mathrm{th}}\bar n(t)
\end{equation}
as in the thermal-loss model introduced above. In $M_2$, this constraint is relaxed and the two rates are reconstructed independently
using
\begin{align}
\kappa_{\downarrow,M_2}(t)
&=
\kappa_{\downarrow,0} +
\alpha_{\downarrow}e^{-t/\tau}\cos(\omega t)
\nonumber\\
\kappa_{\uparrow,M_2}(t)
&=
\kappa_{\uparrow,0} +
\alpha_{\uparrow}e^{-t/\tau}\cos(\omega t)
\label{eq:b2_M2_rates}
\end{align}
with the six fitted parameters
$(\kappa_{\downarrow,0},\kappa_{\uparrow,0},\alpha_{\downarrow},\alpha_{\uparrow},\tau,\omega)$.

For this phase-insensitive generator, local CP-divisibility requires
both instantaneous rates to be non-negative. Equivalently, the
smallest eigenvalue of the local Gaussian CP matrix is
\begin{equation}
\lambda_{\min}
\left[M_{\mathrm{CP}}^{M_2}(t)\right]
= \min \left\{
\kappa_{\downarrow,M_2}(t),
\kappa_{\uparrow,M_2}(t)
\right\}
\label{eq:b2_M2_witness}
\end{equation}
Model $M_2$ does not impose these pointwise positivity conditions;
instead, the accumulated Gaussian channel is constrained to satisfy
the finite-time CP condition in Eq.\eqref{eq:gaussian-cp}. It can therefore represent a locally non-CP-divisible effective dynamics while remaining physical
at every observation time. For the thermal subfamily corresponding to
$M_1$, one has
$\kappa_{\downarrow}-\kappa_{\uparrow}=\kappa_{\mathrm{th}}$
and the above witness reduces to
$\lambda_{\min}[M_{\mathrm{CP}}(t)]
=\kappa_{\mathrm{th}}\bar n(t)$, consistently with Eq.(\ref{eq:b2_thermal_witness}).
Crucially, model $M_3$ is not modified for this benchmark.  We use exactly the resonant one-pseudomode Markovian embedding introduced in
Benchmark 1 with fitted parameters
$(g,\gamma_m)$.  The auxiliary mode and its Markovian reservoir are in vacuum, exactly as in Benchmark 1; no thermal
pseudomode or additional thermal parameter is introduced.  Moreover,
the reference data in the present benchmark contain no auxiliary mode:
$M_3$ is included only as a competing physical hypothesis for the same
homodyne data. Since $M_3$ is a quantum-limited attenuation model, it cannot in general reproduce the excess diffusion generated by
a nonzero thermal occupation.
\begin{table}[t]
\centering
\caption{Model comparison for Benchmark 2. 
The data are generated from a nonstationary thermal excitation process. 
The table shows that the temporal structure in the data is explained without
requiring a non-CP-divisible reduced generator.}
\label{tab:b2_thermal}
\setlength{\tabcolsep}{11pt}      
\renewcommand{\arraystretch}{1.9} 
\begin{tabular}{|c c c c c|}
\hline
Model & $\nu_M$ & $\Delta\mathrm{AIC}$ & $\Delta\mathrm{BIC}$ & $\epsilon^{\rm test}_M$ \\
\hline
$M_0$ & 1 & 65.2 & 31.8 & 1.37734 \\
$M_1$ & 4 & 0.0  & 0.0  & 1.37729 \\
$M_2$ & 6 & 1.4  & 23.7 & 1.37729 \\
$M_3$ & 2 & 26738.2  & 26715.9 & 1.40504 \\
\hline
\end{tabular}
\end{table}
\subsubsection{Fit results and uncertainty assessment}
We use the same single-mode squeezed-vacuum probe
as in Benchmark 1, with $r=0.80$, and the numerical
protocol detailed in Appendix \ref{app:numerical_details}: $T=10$, $K=80$,
four calibrated homodyne phases, $N_{\rm shot}=2000$
samples per time and phase, and an 80/20 training/test split.

The result in Table \ref{tab:b2_thermal} shows that the time-dependent CP-divisible model $M_1$ is the best model by both
information criteria.  The stationary model $M_0$ is disfavored because
it cannot resolve the temporal heating profile.  The more general model
$M_2$ obtains a slightly lower training negative log-likelihood, as
expected from its additional freedom, but the improvement is too small
to compensate for its two additional parameters.  Importantly, its
fitted local witness remains positive:
\begin{equation}
\min_t\lambda_{\min}[M_{\rm CP}^{M_2}(t)]
\simeq 0.0302>0
\label{eq:b2_M2_positive}
\end{equation}
Thus, the data do not support a non-CP-divisible interpretation.
The pseudomode model $M_3$ is also strongly disfavored. Its fitted
dynamics remains a quantum-limited vacuum attenuation process and
therefore cannot reproduce the excess diffusion generated by the
time-dependent thermal occupation in the reference model. This is
physically consistent with the fact that the synthetic reference
contains nonstationary thermal excitation rather than vacuum
pseudomode memory.

\begin{figure}
\centering
\includegraphics[width=\columnwidth]{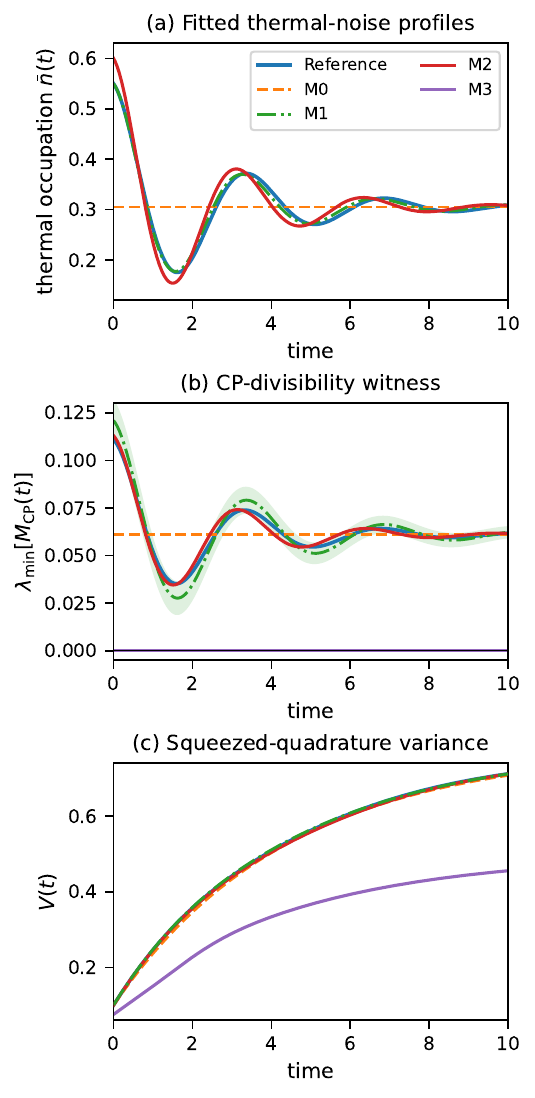}
\caption{Benchmark 2: reconstruction of nonstationary thermal
excitation. (a) Reference thermal occupation and fitted model classes.
The reference profile is time dependent but remains positive. For $M_2$, the plotted thermal occupation is the effective quantity
$\bar n_{\rm eff}(t)
=\kappa_{\uparrow}(t)/
[\kappa_{\downarrow}(t)-\kappa_{\uparrow}(t)]$.
Model $M_3$ is the same vacuum-pseudomode model as in Benchmark 1 and
therefore has no thermal-occupation parameter. (b) Smallest eigenvalue
of the local Gaussian CP matrix.  The witness remains non-negative for
the selected model $M_1$; the light green band shows its 95\% pointwise
bootstrap uncertainty. (c) Squeezed-quadrature variance $V_q(t)$ obtained
from the corresponding fitted dynamics. The $M_1$ and $M_2$ variance curves nearly overlap the
reference.}
\label{fig:b2_thermal}
\end{figure}

To quantify finite-sampling uncertainty, we perform
a parametric bootstrap with $N_{\rm boot}=300$.
Synthetic training data sets are generated from the
fitted $M_1$ channel using the original measurement
settings and sample size. Model $M_1$ is independently
refitted to each realization, and the resulting
distribution is used to construct confidence intervals
for the minimum local CP-divisibility witness.  The resulting 95\% bootstrap interval for the minimum
local CP witness is
\begin{equation}
\min_t\lambda_{\min}[M_{\rm CP}^{M_1}(t)]
\in [0.0242,0.0413]
\label{eq:b2_bootstrap}
\end{equation}
which remains entirely above zero.  The benchmark results
are consistent with the selected CP-divisible model.

Figure \ref{fig:b2_thermal} uses the same three-panel structure as
Benchmark 1, but here the panels emphasize the distinction between
nonstationary thermal excitation and genuinely non-CP-divisible
dynamics. Panel (a) compares the reference thermal occupation with the
profiles reconstructed by the fitted models. The stationary model
$M_0$ can reproduce only an effective average heating level, whereas
$M_1$ follows the temporal modulation of the bath occupation while
remaining within the CP-divisible thermal-loss sector. For $M_2$, we
plot the effective occupation
$\bar n_{\rm eff}(t)
=\kappa_{\uparrow}(t)/
[\kappa_{\downarrow}(t)-\kappa_{\uparrow}(t)]$
wherever the denominator is nonzero. This quantity provides a direct
way of comparing the more general time-local model with the thermal
description, even though in $M_2$ the upward and downward rates are
fitted independently. Model $M_3$ is not shown in panel (a), because it
is the same vacuum-pseudomode model used in Benchmark 1 and therefore
does not possess an independent thermal-occupation parameter.

Panel (b) shows the corresponding local CP-divisibility witness. The
reference process and the selected model $M_1$ remain strictly
non-negative throughout the observation window, demonstrating that the
pronounced time dependence in panel (a) does not by itself imply
non-Markovianity in the CP-divisibility sense. The light green region denotes
the 95\% pointwise bootstrap uncertainty for $M_1$ and remains above
zero, showing that this conclusion is robust against finite-sampling
fluctuations. The fitted $M_2$ trajectory also remains in the
CP-divisible region, so its additional freedom to describe negative
instantaneous rates is not required by the data. For $M_3$, the smallest local CP eigenvalue vanishes
provided that $\kappa_3^{\rm eff}(t)\geq0$; this zero
value reflects saturation of the local CP bound rather
than the absence of dissipative evolution.

Panel (c) translates the reconstructed dynamics into the variance of the initially squeezed $q$-quadrature. More precisely, the plotted quantity is $V_q(t)$, denoted simply by $V(t)$ in the figure, with $V_q(0)=e^{-2r}/2$. For the reference dynamics and $M_1$, the covariance is propagated using Eq.\eqref{eq:second-moment} with the thermal-loss drift and diffusion terms in Eqs.\eqref{photon-loss} and \eqref{eq:D_total_thermal_revised}, where the time dependence enters through the reconstructed bath occupation $\bar n(t)$. Model $M_0$ uses the same thermal-loss evolution but with a constant fitted occupation. For $M_2$, the decay and excitation rates $\kappa_{\downarrow}(t)$ and $\kappa_{\uparrow}(t)$ are reconstructed independently according to Eq.\eqref{eq:b2_M2_rates}, and the corresponding covariance evolution is obtained from these two rates without imposing the thermal-bath relation in Eq.\eqref{eq:thermal_bath}. In contrast, $M_3$ is the vacuum-pseudomode model of Benchmark 1 and therefore describes a quantum-limited attenuation channel; its variance is obtained from the pseudomode transmissivity in Eq.\eqref{eq:b1_M3_eta} using the covariance transformation in Eq.\eqref{eq:covariance_evolution}. The reference dynamics and $M_1$ show the expected gradual increase of $V(t)$, corresponding to degradation of the initial squeezing by thermal excitation, together with a weak modulation inherited from the time-dependent bath occupation. Unlike Benchmark 1, the local rates remain non-negative and no transient recovery of the squeezed-quadrature variance occurs. Thus, the behavior in panel (c) is consistent with the CP-divisibility witness in panel (b): the observed time-dependent heating is explained by a nonstationary but CP-divisible thermal process, without requiring a non-CP-divisible reduced dynamics.

\subsection{Benchmark 3: resonant two-pseudomode memory}
Benchmarks 1 and 2 test, respectively, a phenomenological non-CP-divisible time-local loss process and a nonstationary but CP-divisible thermal process. We now consider a complementary microscopic benchmark in which the synthetic homodyne data are generated directly from the pseudomode embedding introduced in Eq.\eqref{eq:embedded_hamiltonian}. The purpose of this benchmark is to test whether the reconstruction can identify an explicit physical memory mechanism when the reduced photonic dynamics is non-CP-divisible and the embedded model $M_3$ is the correct description.

Physically, benchmark 3 represents a structured vacuum-loss environment
with two hidden resonant modes. A possible realization is an observed
photonic mode coupled coherently to two weakly damped auxiliary resonances,
representing, for example, parasitic cavity modes, spectral side modes, or
delay-line modes of the photonic circuit. Each auxiliary mode is itself
coupled to an independent Markovian vacuum reservoir. Photons can therefore
leave the observed mode, remain temporarily stored in the auxiliary modes,
and subsequently return before being irreversibly lost to the external
reservoirs. After eliminating the auxiliary modes, this coherent
storage-and-return mechanism appears as a time-correlated vacuum-loss channel
with a two-exponential memory kernel. Thus, benchmark 3 provides a microscopic
realization of non-Markovian photon loss rather than thermal excitation or
classical technical noise.

\subsubsection{Candidate models}
We consider one observed photonic mode $a$ coupled to two auxiliary bosonic memory modes $b_1$ and $b_2$. The three modes are taken to be resonant. The interaction in Eq. \eqref{eq:embedded_hamiltonian} is written in the rotating-wave approximation (RWA), retaining only the excitation-conserving terms $a^\dagger b_\ell$ and $a b_\ell^\dagger$ while neglecting the counter-rotating terms $a b_\ell$ and $a^\dagger b_\ell^\dagger$ \cite{Breuer2007}.
In a frame rotating at the common resonance frequency,\footnote{A rotating frame is obtained by a unitary
transformation that removes the free oscillation at a chosen
reference frequency, leaving detunings and interaction terms
in the transformed Hamiltonian. For the rotating-wave
approximation (RWA) used here, we assume
$g_\ell,\gamma_\ell\ll\omega_0$, where $\omega_0$ is the
common resonance frequency.} the free-frequency terms vanish and the detunings are zero, so the embedded Hamiltonian becomes \cite{Gardiner2004,Walls2008}
\begin{equation}
H_{\rm emb}^{(2)} =
g_1\left(a^\dagger b_1+a b_1^\dagger\right)
+g_2\left(a^\dagger b_2+a b_2^\dagger\right)
\label{eq:b3_Hemb}
\end{equation}
where the coupling constants $g_1$ and $g_2$ are taken to be real. Their phases can be absorbed into independent phase definitions of $b_1$ and $b_2$.

To match the embedding of Eqs.\eqref{eq:embedded_hamiltonian}-\eqref{eq:pseudomode_memory_kernel} exactly, the observed mode $a$ is not coupled directly to an additional Markovian reservoir. Instead, environmental damping acts only on the two pseudomodes. In this benchmark we denote the pseudomode decay rates by $\gamma_1$ and $\gamma_2$. After tracing out the two memoryless vacuum reservoirs coupled to $b_1$ and $b_2$, the joint density operator of the explicit $a+b_1+b_2$ subsystem obeys
\begin{equation}
\frac{d\varrho}{dt}
=-i\left[H_{\rm emb}^{(2)},\varrho\right]
+\gamma_1\mathcal D[b_1]\varrho
+\gamma_2\mathcal D[b_2]\varrho 
\label{eq:b3_master}
\end{equation}
Thus, the enlarged $a+b_1+b_2$ subsystem evolves according to a standard CP Markovian Lindblad equation even when the reduced dynamics of the observed mode $a$ is not CP divisible.

The corresponding Heisenberg-Langevin equations are \cite{Breuer2007}
\begin{align}
\dot a(t)
&=-i g_1 b_1(t)-i g_2 b_2(t)
\label{eq:b3_aLangevin}\\
\dot b_1(t)
&=-\frac{\gamma_1}{2}b_1(t)-i g_1 a(t)
+\sqrt{\gamma_1}\,b_{1,{\rm in}}(t)
\label{eq:b3_b1Langevin}\\
\dot b_2(t)
&=-\frac{\gamma_2}{2}b_2(t)-i g_2 a(t)
+\sqrt{\gamma_2}\,b_{2,{\rm in}}(t)
\label{eq:b3_b2Langevin}
\end{align}
Here $a(t)$ is the annihilation operator of the observed photonic mode and $b_1(t),b_2(t)$ are the two explicit pseudomode operators retained in the enlarged system. In contrast, $b_{1,{\rm in}}(t)$ and $b_{2,{\rm in}}(t)$ are the input-field operators of the independent Markovian vacuum reservoirs coupled to the pseudomodes. The terms $-\gamma_\ell b_\ell/2$ describe pseudomode damping and the terms $\sqrt{\gamma_\ell}\,b_{\ell,{\rm in}}$ provide the corresponding vacuum Langevin noise required by the damping channel.

For $c_{\mu,{\rm in}}\in\{b_{1,{\rm in}},b_{2,{\rm in}}\}$, the independent vacuum inputs satisfy
\begin{align}
\left\langle c_{\mu,{\rm in}}(t)c_{\nu,{\rm in}}^\dagger(s)\right\rangle
&=\delta_{\mu\nu}\delta(t-s)
\label{eq:b3_vac1}\\
\left\langle c_{\mu,{\rm in}}^\dagger(t)c_{\nu,{\rm in}}(s)\right\rangle
&=0
\label{eq:b3_vac2}
\end{align}
The homogeneous part of Eqs.\eqref{eq:b3_aLangevin}-\eqref{eq:b3_b2Langevin} is generated by
\begin{equation}
P_3=
\begin{pmatrix}
0 & -i g_1 & -i g_2\\
-i g_1 & -\gamma_1/2 & 0\\
-i g_2 & 0 & -\gamma_2/2
\end{pmatrix}
\label{eq:b3_P3}
\end{equation}
The absence of diagonal frequency offsets is the zero-detuning resonance condition, while the zero in the $(1,1)$ entry also reflects the absence of a direct Markovian loss channel on $a$.

The two-memory kernel follows by formally integrating the pseudomode equations. For $\ell=1,2$, we have
\begin{equation}
\begin{aligned}
b_\ell(t)
=&e^{-\gamma_\ell t/2}b_\ell(0)
-i g_\ell\int_0^t e^{-\gamma_\ell(t-s)/2}a(s)\,ds\\
&+\sqrt{\gamma_\ell}\int_0^t e^{-\gamma_\ell(t-s)/2}b_{\ell,{\rm in}}(s)\,ds 
\label{eq:b3_bsolution}
\end{aligned}
\end{equation}
Substituting Eq.\eqref{eq:b3_bsolution} into Eq.\eqref{eq:b3_aLangevin} gives
\begin{align}
\dot a(t)
={}&-\int_0^t K(t-s)a(s)\,ds
-i\sum_{\ell=1}^{2}g_\ell e^{-\gamma_\ell t/2}b_\ell(0)
\nonumber\\
&-i\sum_{\ell=1}^{2}g_\ell\sqrt{\gamma_\ell}
\int_0^t e^{-\gamma_\ell(t-s)/2}b_{\ell,{\rm in}}(s)\,ds
\label{eq:b3_nonlocal}
\end{align}
where the resonant memory kernel is
\begin{equation}
K(t)=g_1^2e^{-\gamma_1 t/2}+g_2^2e^{-\gamma_2 t/2},
\qquad t\ge0
\label{eq:b3_kernel}
\end{equation}
Eq.\eqref{eq:b3_kernel} is the two-pseudomode specialization of Eq.\eqref{eq:pseudomode_memory_kernel}, with the same decay-rate notation
$\gamma_\ell$. The two exponentials introduce the
memory timescales $2/\gamma_1$ and $2/\gamma_2$.

Let $\chi_3(t)$ denote the coefficient multiplying $a(0)$ in the solution for $a(t)$. Since the pseudomode initial operators and vacuum input fields in Eq.\eqref{eq:b3_nonlocal} do not contain $a(0)$, this coefficient obeys
\begin{equation}
\dot\chi_3(t)
=-\int_0^t K(t-s)\chi_3(s)\,ds,
\qquad \chi_3(0)=1
\label{eq:b3_chi_int}
\end{equation}
Equivalently
\begin{equation}
\chi_3(t)=\left[e^{P_3t}\right]_{11}
\label{eq:b3_chi_exp}
\end{equation}
Therefore, $\chi_3(t)$ is determined by the three eigenmodes of
$P_3$, giving a three-pole response that is more structured than the
single damped-oscillatory parameterization used for $M_1$ and $M_2$. The transfer amplitude $\chi_3(t)$ has isolated zeros,
at which the reduced time-local generator is singular.
Fitting therefore uses the embedded dynamics and the
finite-time channel, which remain well defined through
these points; generator reconstruction and its diagnostics
are restricted to nonsingular intervals. By contrast,
a bounded damped-cosine rate yields strictly positive
transmissivity at every finite time and cannot reproduce
these exact zeros.

The two pseudomodes and their two Markovian reservoirs are initially in vacuum. Because the embedding is passive and number preserving, the reduced channel of the observed mode is therefore a quantum-limited attenuation channel with transmissivity $\eta$ that obeys Eq.\eqref{eq:b1_M3_eta}.
For the squeezed-vacuum probe used in the previous benchmarks, the initial variances of the $q$- and $p$-quadratures of the squeezed-vacuum input state $
V_q(0)=\frac{1}{2}e^{-2r}$ and $
V_p(0)=\frac{1}{2}e^{2r}$
and they evolve as Eq.\eqref{eq:covariance_evolution}.
Away from isolated zeros of $\chi_3(t)$, the corresponding time-local reduced loss rate is defined in Eq.\eqref{eq:b1_M3_effective_rate}. Generator-level diagnostics are evaluated only away
from transfer-amplitude zeros; evolution through
these zeros is described using the finite-time
Gaussian channel.

For a vacuum attenuation channel, the two eigenvalues of the local Gaussian CP matrix are $0$ and $\kappa_3^{\rm eff}(t)$. Hence
\begin{equation}
\lambda_{\min}[M_{\rm CP}(t)]
=\min\left\{0,\kappa_3^{\rm eff}(t)\right\}
\label{eq:b3_witness}
\end{equation}
Whenever $\eta(t)$ increases, $\dot\eta(t)>0$ and $\kappa_3^{\rm eff}(t)<0$. A transmissivity revival therefore identifies non-CP-divisible reduced dynamics, while the enlarged dynamics in Eq.\eqref{eq:b3_master} remains CP and Markovian.

\subsubsection{Fit results and uncertainty assessment}
We use the same numerical protocol as in the previous benchmarks. The reference parameters are
\begin{equation}
g_1=0.60,
\quad g_2=0.20,
\quad \gamma_1=0.12,
\quad \gamma_2=0.45
\label{eq:b3_refpars}
\end{equation}
The first pseudomode is relatively long-lived,
with a memory lifetime $1/\gamma_1 \simeq 8.3$,
and provides the dominant coherent storage-and-return
pathway, whereas the second pseudomode introduces
a shorter memory component. Their combined effect produces two attenuation revivals over the observation window.

The candidate models retain the same hierarchy used above. Model $M_0$ is stationary pure loss with one fitted rate. Models $M_1$ and $M_2$ use the same four-parameter damped-cosine rate in Eq.\eqref{eq:b1_common_family}.
Model $M_1$ imposes $\kappa_{M_1}(t)\ge0$ throughout the observation window. Model $M_2$ allows a signed instantaneous rate but requires the accumulated attenuation channel to remain completely positive. Model $M_3$ is the resonant two-pseudomode embedding in Eq.\eqref{eq:b3_master} and has four fitted real parameters,
$(g_1,g_2,\gamma_1,\gamma_2)$.

\begin{table}[t]
\centering
\caption{Model comparison for Benchmark 3. The reference homodyne data are generated by the resonant two-pseudomode embedding without a direct Markovian loss channel on the observed mode. Relative AIC and BIC are reported with respect to the best model, while $\epsilon_M^{(\mathrm{test})}$ is the held-out negative log-likelihood per test sample.}
\label{tab:b3}
\setlength{\tabcolsep}{11pt}      
\renewcommand{\arraystretch}{1.9} 
\begin{tabular}{|c c c c c|}
\hline
Model & $\nu_M$ & $\Delta\mathrm{AIC}$ & $\Delta\mathrm{BIC}$ & $\epsilon^{\rm test}_M$ \\
\hline
$M_0$ & 1 & 33719.0 & 33685.6 & 1.26134 \\
$M_1$ & 4 & 21765.0  & 21765.0  & 1.24929 \\
$M_2$ & 4 & 18204.4  & 18204.4 & 1.24599 \\
$M_3$ & 4 & 0.0  & 0.0 & 1.22806 \\
\hline
\end{tabular}
\end{table}

Table \ref{tab:b3} exhibits the intended hierarchy of model performance. The stationary model $M_0$ cannot reproduce the coherent memory revivals. Model $M_1$ follows part of the temporal variation, but its pointwise positivity constraint forbids information return. Model $M_2$ performs substantially better than $M_1$ despite having the same number of fitted parameters. The best-fit solution for $M_2$, obtained by minimizing
the training negative log-likelihood, has an instantaneous
loss rate that becomes negative
\begin{equation}
\min_{t\in[0,T]}\kappa_{M_2}(t)\simeq -0.470
\label{eq:b3_m2min}
\end{equation}
This negative reconstructed rate is consistent with non-CP-divisible reduced dynamics, although the model comparison alone does not establish their statistical necessity. Nevertheless, the single damped-cosine rate of $M_2$ cannot reproduce the complete three-pole response produced by the two pseudomodes. Model $M_3$ is selected decisively by both information criteria and gives the best held-out predictive score. Since $M_1$, $M_2$, and $M_3$ all contain four fitted parameters, this preference reflects the two-pseudomode memory structure rather than a larger parameter count.

To quantify the statistical uncertainty of the selected memory model, we perform a parametric bootstrap with $N_{\rm boot}=300$. Each bootstrap training set is generated from the fitted $M_3$ channel using the same measurement times, homodyne phases, and number of training shots as the original benchmark, and $M_3$ is refitted independently to every bootstrap realization. 
At exact resonance, $\chi_3(t)$ has isolated zeros
near the coherent-exchange minima, where the
effective loss rate $\kappa_3^{\rm eff}(t)$ becomes
singular, although the finite-time transmissivity
remains regular. We therefore evaluate the local
CP-divisibility witness on the regular interval
$I=[3.3,4.8]$, away from these singularities.
Since the minimum witness value depends on the
chosen interval boundary, we instead assess
whether its bootstrap upper confidence bound
remains negative throughout this interval.
For $N_{\rm boot}=300$ parametric bootstrap
refits of the selected $M_3$ model, the maximum
pointwise 97.5\% upper bound is $-0.139$.
The reconstructed witness therefore remains
negative across the considered interval within
the pointwise bootstrap uncertainty, supporting
non-CP-divisible reduced dynamics conditional
on the fitted model.

\begin{figure}
\centering
\includegraphics[width=0.92\linewidth]{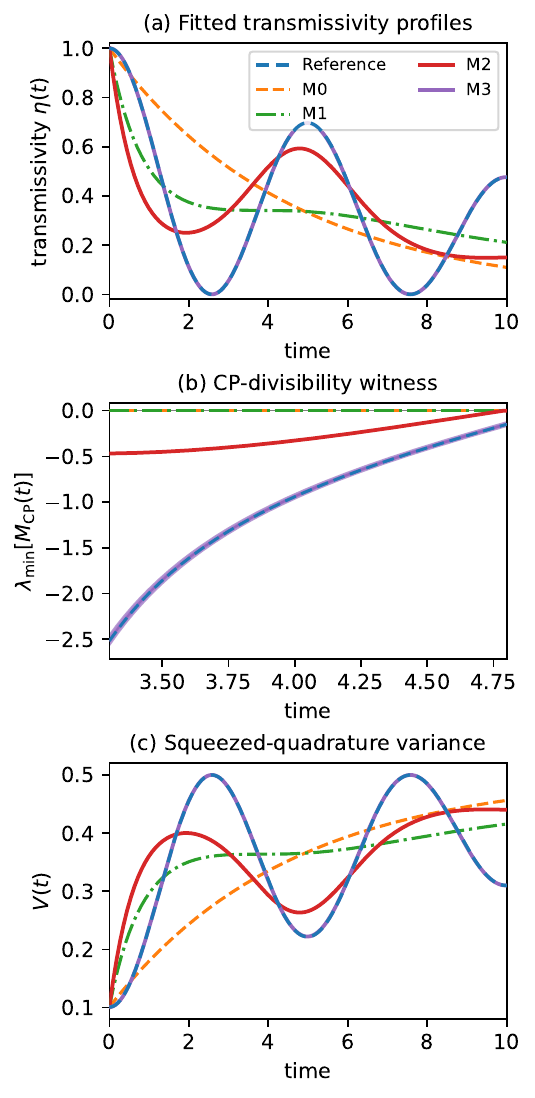}
\caption{Benchmark 3: resonant two-pseudomode memory without direct Markovian loss of the observed mode. (a) Reference and fitted finite-time transmissivities. The blue Reference curve is the ground-truth two-pseudomode model, while the purple $M_3$ curve is the fitted reconstruction within the same model class. The transmissivity is used instead of the full-time effective loss rate because the latter has isolated representation poles when $\chi_3(t)$ crosses zero. 
(b) Smallest eigenvalue of the local Gaussian
CP matrix on the regular interval $I=[3.3,4.8]$,
chosen away from singularities of the effective
time-local generator. The shaded region represents
the 95\% pointwise bootstrap interval for $M_3$.
The upper bound remains negative throughout
the displayed interval, supporting the reconstructed
violation of CP-divisibility.
 (c) Squeezed-quadrature variance $V(t)\equiv V_q(t)$. The quadrature recovery follows the transmissivity revival and provides an observable signature of pseudomode-mediated information return.}
\label{fig:b3}
\end{figure}

Figure \ref{fig:b3} illustrates how the two pseudomodes generate and store memory in the reduced photonic dynamics. In panel (a), the effective transmissivity of the observed mode, defined in Eq.\eqref{eq:b1_M3_eta}, exhibits
pronounced revivals due to coherent excitation exchange with the two damped pseudomodes. The fitted $M_3$ model
closely reproduces the reference dynamics. We plot the transmissivity rather than the effective loss rate
$\kappa_3^{\rm eff}(t)$ because the latter is singular at isolated zeros of $\chi_3(t)$, whereas the transmissivity remains well defined. Panel (b) shows that these revivals are accompanied by negative values of the local CP-divisibility witness, demonstrating information backflow in the reduced mode while the enlarged pseudomode dynamics remains Markovian. 
Panel (c) shows the corresponding evolution of the squeezed-quadrature
variance. As the transmissivity decreases, the initially squeezed variance
increases toward the vacuum value, reflecting degradation of the squeezing.
When the transmissivity subsequently revives because excitation returns
from the pseudomodes to the observed mode, the quadrature variance decreases
again and part of the squeezing is recovered. The close agreement between
the reference and fitted $M_3$ curves therefore shows that the reconstructed
pseudomode model reproduces not only the underlying memory dynamics but also
its directly observable signature in the homodyne variance.

Consequently, the three benchmarks demonstrate discrimination
between CP-divisible and non-CP-divisible dynamics within
the tested model families. The preference for a pseudomode
description does not uniquely identify the underlying
environment, since different environmental realizations
can produce identical single-time homodyne statistics.

\section{Conclusion}
\label{sec:conclusion}
The main contribution of this work is a physically interpretable framework for reconstructing non-Markovian Gaussian noise in CV photonic quantum systems. 
The theoretical framework connects reconstructed Gaussian
generators to physical noise mechanisms, including photon
loss, thermal excitation, correlated dissipation, and
coupling to auxiliary memory modes, while also discussing
effective descriptions of technical fluctuations.
The numerical benchmarks demonstrate the distinction
between CP-divisible and non-CP-divisible dynamics
for single-mode attenuation and thermal-noise processes.
A key advantage of the framework is its model hierarchy, which ranges from static Markovian noise to time-dependent Markovian dynamics, Lindblad-like non-Markovian generators, and pseudomode-embedded Markovian models. This hierarchy provides a systematic framework for comparing CP-divisible and non-CP-divisible descriptions of the observed dynamics. Non-Markovianity in the CP-divisibility sense is diagnosed
by negative eigenvalues of $M_{\rm CP}(t)$. Covariance
revivals and improved predictive performance provide
supporting evidence in the present benchmarks, but are
not general stand-alone criteria for non-Markovianity.

The proposed reconstruction procedure combines Gaussian moment propagation,
likelihood-based parameter estimation, physicality constraints, and
model-selection criteria. 
In the numerical benchmarks, this combination distinguishes genuinely
non-CP-divisible reduced dynamics from ordinary time-dependent Gaussian
noise across three attenuation-related physical scenarios: time-dependent
photon loss, nonstationary thermal loss with excitation, and a structured
non-Markovian vacuum-loss environment with two hidden resonant modes.
The thermal benchmark shows that time-dependent diffusion and heating need
not imply non-Markovianity, whereas the two-pseudomode benchmark provides
a microscopic interpretation of genuine memory in terms of coherent
storage and return of excitation from hidden environmental modes.
As a result, the method provides both a quantitative reconstruction of the
unknown noise dynamics and a physical interpretation of its possible
microscopic origin.
This is especially useful for CV photonic platforms, where memory effects may
arise from structured reservoirs, spectral side modes, imperfect filtering,
feedback paths, or slowly fluctuating control fields.

The present study is restricted to Gaussian-preserving
dynamics and simulation-based validation. Although
the theoretical framework includes correlated dissipation
and effective descriptions of technical fluctuations,
the numerical benchmarks focus on single-mode attenuation,
thermal excitation, and pseudomode-mediated memory.
Multiplicative fluctuations, such as pump-phase and
squeezing-strength noise, require additional approximations
and are not explicitly validated here. Future work may
extend the numerical validation to correlated noise,
multimode photonic systems, and experimental homodyne
data, as well as investigate adaptive tomography and
non-Gaussian noise.

\subparagraph{Acknowledgments.}
The author would like to thank Sevag Gharibian and Michael Stefszky for their valuable discussions, insightful comments, and constructive suggestions that contributed to the development of this work. The author acknowledges support from the PhoQuant project, funded by the Bundesministerium für Forschung, Technologie und Raumfahrt (BMFTR).

\subparagraph{Data availability.}
Further details of the numerical implementation, simulation parameters, and validation procedures are available from the corresponding author upon reasonable request.

\appendix
\section{Derivation of the Gaussian propagator and local CP condition}
\label{app:Gaussian-propagator}
Starting from the Gaussian moment relations in Eqs.\eqref{eq:first-moment} and \eqref{eq:second-moment}
\begin{align}
\dot {\bm{d}}(t) &= A(t)\bm d(t)+\bm u(t),\\
\dot {V}(t) &= A(t) V(t)+ V(t)A^T(t)+D(t)
\end{align}
we introduce the fundamental matrix $X(t,s)$, defined as the solution of
\begin{equation}
\frac{d}{dt}X(t,s)=A(t)X(t,s),\qquad X(s,s)=I 
\end{equation}
Multiplying the first-moment equation by $X^{-1}(t,0)$ and integrating from $0$ to $t$ gives
\begin{equation}
\bm d(t)=X(t,0)\bm d(0)+\int_0^t X(t,s)\bm u(s)ds 
\end{equation}
For the covariance matrix, we define
\begin{equation}
\widetilde V(t)=X^{-1}(t,0)V(t)X^{-T}(t,0)
\end{equation}
Differentiating this expression and using the covariance equation cancels the drift terms, leaving
\begin{equation}
\dot{\widetilde V}(t)=X^{-1}(t,0)D(t)X^{-T}(t,0)
\end{equation}
Integrating and multiplying back by $X(t,0)$ and $X^T(t,0)$ gives
\begin{equation}
 V(t)=X(t,0) V(0)X^T(t,0)
+\int_0^t X(t,s)D(s)X^T(t,s) ds 
\end{equation}
Thus, the Gaussian channel can be written in the standard form
\begin{equation}
V(t)=X(t)V(0)X^T(t)+Y(t)
\end{equation}
where
\begin{equation}
Y(t)=\int_0^t X(t,s)D(s)X^T(t,s) ds 
\end{equation}
The CP condition for this Gaussian channel can be defined \cite{Ernesto2026,Mirkhalaf2024}
\begin{equation}
Y(t)+\frac{i}{2}\Omega-\frac{i}{2}X(t)\Omega X^T(t)\ge 0 
\end{equation}
For an infinitesimal map from $t$ to $t+dt$,
assuming sufficiently smooth drift and diffusion
matrices, one has
\begin{align}
&X(t+dt,t)=I+A(t)dt+O(dt^2)\nonumber\\ &Y(t+dt,t)=D(t)dt+O(dt^2)
\end{align}
Substituting these expressions into the Gaussian CP condition and keeping only first-order terms in $dt$ gives
\begin{equation}
D(t) - \frac{i}{2}
\left[ A(t)\Omega+\Omega A^T(t) \right]
\geq 0 
\end{equation}
This is the infinitesimal CP condition used to test CP-divisibility of the Gaussian dynamics.

\section{Numerical implementation and synthetic-data generation}
\label{app:numerical_details}
All numerical simulations are performed in dimensionless units, with the vacuum covariance normalized as $I_2/2$. Time is measured in units of an arbitrary reference scale $t_0$, so that $t=t_{\rm phys}/t_0$. Accordingly, the dimensionless rates and frequencies reported below
correspond to physical rates and frequencies measured in units of
$t_0^{-1}$. The continuous interval $t\in[0,T]$ is sampled at discrete measurement times $t_k$, $k=1,\ldots,K$, at which the homodyne data are generated and the reconstructed dynamics are evaluated. 
The total simulation time is $T=10$ and the interval
$[0,T]$ is discretized into $K=80$ equally spaced measurement times. These choices provide sufficient temporal
resolution to resolve the benchmark memory times and
oscillation frequencies. Their conversion to laboratory
units depends on the platform-dependent reference time
$t_0$.
For each time, homodyne data are generated for the phases
\begin{equation}
\theta \in \{0,\pi/4,\pi/2,3\pi/4\}
\label{eq:phases}
\end{equation}
For every time, phase and input probe state, we use $\Nshot=2000$ synthetic homodyne samples. 
Each homodyne shot samples a separately prepared evolution,
with the system reinitialized and the memory modes and
environment restored to the same reference state.
Different measurement times therefore correspond to
independent preparations, not repeated destructive
measurements of one trajectory. The product likelihood
and random-shot train/test split assume that resetting
and sufficient waiting between shots suppress residual
environmental and detector correlations, with negligible
drift during acquisition. If these conditions fail, a
likelihood accounting for correlations between shots and
block-based splitting and resampling are required.

The data are split into training and test sets using an $80/20$ split. The training data are used to estimate the parameters of each candidate model, whereas the test data are kept separate from the fitting procedure
and are used only to evaluate how well the fitted model predicts independent homodyne samples \cite{mogilevtsev2013cross,tiunov2020experimental}. This separation helps determine whether the improved training fit of a
more flexible model reflects genuine dynamical structure or merely
overfitting to statistical noise in the training data.

For an input probe state labelled by $\alpha$, the real-valued synthetic homodyne outcomes are generated as
\begin{equation}
x^{(\alpha)}_{k\ell s} \sim
\mathcal{N} \left(
\mathbf{e}_\ell^{T}\mathbf{d}^{\rm true}_{\alpha}(t_k),
\,\mathbf{e}_\ell^{T}V^{\rm true}_{\alpha}(t_k)\mathbf{e}_\ell
\right)
\label{eq:validation_data_generation}
\end{equation}
where $\mathcal{N}$ denotes a normal distribution, $t_k$ is the measurement time, $\ell=1,\ldots,L$ labels the homodyne phase $\theta_\ell$ in Eq.\eqref{eq:phases}, $\mathbf{e}_{\ell}$ is the homodyne direction and $s$ labels the shot. The reference moments $\mathbf{d}_{\alpha}^{\mathrm{true}}(t)$ and $V_{\alpha}^{\mathrm{true}}(t)$ are obtained by solving the Gaussian moment equations of the preceding sections.
The reference dynamics is used consistently for data generation, model fitting and plotting. 

The following steps clarify how the reference dynamics, synthetic homodyne data, fitted models, and plotted curves are connected. First, the exact benchmark model defines $\mathbf{d}_\alpha^{\rm true}(t)$ and $V_\alpha^{\rm true}(t)$ through the Gaussian
moment equations. Synthetic homodyne samples are then drawn from the corresponding Gaussian distributions in Eq.\eqref{eq:validation_data_generation}. The candidate models
$M_0$-$M_3$ are fitted only to the training data, while the test data are
used only for predictive comparison. The curves shown in the figures of each benchmark are
therefore not independent assumptions: they are the reference or fitted
time-local coefficients and the corresponding propagated observables.

The numerical values are constrained benchmark parameters chosen to represent moderate optical loss or diffusion over the full observation window, together with finite memory times comparable to a few circuit time units. The memory lifetimes $\tau_c=2.5$-$3.5$ and oscillation frequencies $\omega_c=1.8$-$2.2$ describe damped memory kernels of the type produced by resonator modes, delay-line feedback, finite-bandwidth technical noise, or pseudomode embeddings \cite{Breuer2007,Garraway1997,Pleasance2020}. The oscillatory components represent a moderate memory regime in which the
instantaneous time-local generator can become temporarily non-CP-divisible,
while the accumulated finite-time Gaussian channel remains CP.

In all benchmarks, $\nu_M$ denotes the number of real scalar parameters that
are varied during the fit. Parameters fixed by the benchmark design, such as the sampled time
grid, homodyne angles, and input-state squeezing, are not counted in $\nu_M$.
For the numerical benchmarks, each candidate model is described by a
low-dimensional physically motivated parameter set $\Theta_M$.
Accordingly, the parameters are obtained by minimizing the training
negative log-likelihood in Eq.\eqref{eq:negative-log-likelihood}, subject to the model-specific
physicality constraints described below. The regularized estimator in
Eq.\eqref{eq:regularized-estimator} is intended for more general reconstructions in which
$A(t)$, $D(t)$, and $\bm{u}(t)$ are represented by flexible basis expansions in Eqs.\eqref{eq:A-basis}-\eqref{eq:u-basis-expansion}.

The AIC and BIC criteria defined in
Eqs.\eqref{eq:AIC} and \eqref{eq:BIC} are evaluated using the unregularized training negative
log-likelihood,
$\mathcal{J}_{M}^{\mathrm{tr}}(\Theta_M)$ and $N$ is replaced by the number of independent training homodyne samples $N_{\rm tr}$.
The test predictive score $\epsilon_M^{\mathrm{test}}$ is evaluated separately using Eq.\eqref{eq:test-nll}.
To compare the candidate models independently of the absolute scale of the information criteria, we subtract
the smallest AIC and BIC values obtained among the four candidate models as a reference \footnote{We report relative AIC and BIC values because only differences
between candidate models are relevant for model comparison; subtracting
the minimum sets the best-supported model to zero and makes the comparison
more transparent.}
\begin{equation}
\begin{aligned}
&\Delta\AIC_{M}
=\AIC_{M}-\min_{M'}\AIC_{M'}\\
&\Delta\BIC_{M}
=\BIC_{M}-\min_{M'}\BIC_{M'}
\label{eq:delta_AIC}
\end{aligned}
\end{equation}
where $M'$ is a dummy
model index over the full model set 
$\{M_0,M_1,M_2,M_3\}$ used to identify
the best AIC or BIC value.
Lower values of $\Delta\AIC_M$ and $\Delta\BIC_M$ indicate stronger
information-criterion support relative to the best candidate model, while a
lower $\epsilon_M^{\mathrm{test}}$ indicates better test predictive performance.

In the numerical benchmarks, the matrices $A_M(t)$ and $D_M(t)$ are not treated as arbitrary matrix-valued functions.
Instead, each candidate model is described by a low-dimensional, physically motivated parameter set $\Theta_M$. These parameters
determine the corresponding drift and diffusion matrices or, for pseudomode models, the corresponding reduced Gaussian channel from which the moments are propagated. In all three benchmarks, the displacement drift vanishes, $\bm{u}_M(t)=0$.
For the time-dependent phenomenological models $M_1$ and $M_2$, we employ damped oscillatory parameterizations as controlled
low-dimensional ans\"atze. This choice facilitates comparison between models and should not be interpreted as implying either a common microscopic origin or non-Markovianity from oscillatory time dependence alone.

The numerical benchmarks use SciPy routines to minimize
the Gaussian training negative log-likelihood within
prescribed parameter bounds. The stationary model $M_0$
uses the limited-memory Broyden-Fletcher-Goldfarb-Shanno
algorithm with bound constraints (L-BFGS-B). In Benchmarks 1
and 2, $M_1$ and $M_2$ use sequential least squares
programming (SLSQP) with multiple initial guesses, whereas
$M_3$ uses differential evolution followed by L-BFGS-B
refinement. In Benchmark 3, $M_1$, $M_2$, and $M_3$ use
differential evolution with multiple random seeds and local
polishing. 
The numerical optimization employs model-specific parameter
bounds, multiple initial guesses, and prescribed stopping
tolerances, with different random initializations where
applicable.
Physicality constraints are evaluated on separate
grids containing 600, 300, and 1001 points for Benchmarks 1,
2, and 3, respectively, using explicit inequalities or large
objective penalties. These checks provide discrete
approximations to the continuous-time constraints.
Channel propagation uses analytical expressions where
available; the diffusion contribution for $M_2$ in
Benchmark 2 is integrated using an adaptive Runge-Kutta
method of order 5(4), with relative and absolute tolerances
$2\times10^{-8}$ and $10^{-10}$. Pseudomode models propagate
the finite-time channel directly, avoiding division by the
transfer amplitude during fitting. In Benchmark 3, the
time-local witness is evaluated on a restricted interval
away from transfer-amplitude zeros. Independent test data
are used only for predictive evaluation.

\bibliographystyle{unsrt}
\bibliography{refs}

@article{Lindblad1976,
  author  = {Lindblad, Göran},
  title   = {On the generators of quantum dynamical semigroups},
  journal = {\textnormal{\href{https://doi.org/10.1007/BF01608499}{Communications in Mathematical Physics 48, 119 (1976)}}},
  volume  = {},
  number  = {},
  pages   = {},
  year    = {},
  doi     = {10.1007/BF01608499}
}

@article{Gorini1976,
  author  = {Gorini, Vittorio and Kossakowski, Andrzej and Sudarshan, E. C. G.},
  title   = {Completely positive dynamical semigroups of {$N$}-level systems},
  journal = {\textnormal{\href{https://doi.org/10.1063/1.522979}{Journal of Mathematical Physics 17, 821 (1976)}}},
  volume  = {},
  number  = {},
  pages   = {},
  year    = {},
  doi     = {10.1063/1.522979}
}

@article{Breuer2009,
  author  = {Breuer, Heinz-Peter and Laine, Elsi-Mari and Piilo, Jyrki},
  title   = {Measure for the degree of non-{M}arkovian behavior of quantum processes in open systems},
  journal = {\textnormal{\href{https://doi.org/10.1103/PhysRevLett.103.210401}{Physical Review Letters 103, 210401 (2009)}}},
  volume  = {},
  number  = {},
  pages   = {},
  year    = {},
  doi     = {10.1103/PhysRevLett.103.210401}
}

@article{Rivas2014,
  author  = {Rivas, Ángel and Huelga, Susana F. and Plenio, Martin B.},
  title   = {Quantum non-{M}arkovianity: characterization, quantification and detection},
  journal = {\textnormal{\href{https://doi.org/10.1088/0034-4885/77/9/094001}{Reports on Progress in Physics 77, 094001 (2014)}}},
  volume  = {},
  number  = {},
  pages   = {},
  year    = {},
  doi     = {10.1088/0034-4885/77/9/094001}
}

@article{Weedbrook2012,
  author  = {Weedbrook, Christian and others},
  title   = {Gaussian quantum information},
  journal = {\textnormal{\href{https://doi.org/10.1103/RevModPhys.84.621}{Reviews of Modern Physics 84, 621 (2012)}}},
  volume  = {},
  number  = {},
  pages   = {},
  year    = {},
  doi     = {10.1103/RevModPhys.84.621}
}

@book{Serafini2017,
  author    = {Serafini, Alessio},
  title     = {Quantum Continuous Variables: A Primer of Theoretical Methods},
  publisher = {\textnormal{\href{https://doi.org/10.1201/9781315118727}{CRC Press, 2017}}},
  address   = {},
  year      = {},
  doi     = {10.1201/9781315118727}
}

@article{Zhang2019,
  author  = {Zhang, Gong and others},
  title   = {An integrated silicon photonic chip platform for continuous-variable quantum key distribution},
  journal = {\textnormal{\href{https://doi.org/10.1038/s41566-019-0504-5}{Nature Photonics 13, 839–842 (2019)}}},
  volume  = {},
  pages   = {},
  year    = {},
  doi     = {10.1038/s41566-019-0504-5}
}

@article{Usenko2026,
  author  = {Usenko, Vladyslav C and others},
  title   = {Continuous-variable quantum communication},
  journal = {\textnormal{\href{https://doi.org/10.1103/mgj7-t6d3}{Reviews of Modern Physics 98, 015003  (2026)}}},
  volume  = {},
  pages   = {},
  year    = {},
  doi     = {10.1103/mgj7-t6d3}
}

@article{Lvovsky2009,
  author  = {Lvovsky, Alexander and Raymer, Michael G.},
  title   = {Continuous-variable optical quantum-state tomography},
  journal = {\textnormal{\href{https://doi.org/10.1103/RevModPhys.81.299}{Reviews of Modern Physics 81, 299 (2009)}}},
  volume  = {},
  pages   = {},
  year    = {},
  doi     = {10.1103/RevModPhys.81.299}
}

@article{Lloyd1999,
  author  = {Lloyd, Seth and Braunstein, Samuel L.},
  title   = {Quantum Computation over Continuous Variables},
  journal = {\textnormal{\href{https://doi.org/10.1103/PhysRevLett.82.1784}{Physical Review Letters 82, 1784 (1999)}}},
  volume  = {},
  pages   = {},
  year    = {},
  doi     = {10.1103/PhysRevLett.82.1784}
}

@article{Braunstein2005,
  author  = { Braunstein, Samuel L. and van Loock, Peter},
  title   = {Quantum information with continuous variables},
  journal = {\textnormal{\href{https://doi.org/10.1103/RevModPhys.77.513}{Reviews of Modern Physics 77, 513 (2005)}}},
  volume  = {},
  pages   = {},
  year    = {},
  doi     = {10.1103/RevModPhys.77.513}
}

@article{Adesso2014,
  author  = {Adesso, Gerardo and Ragy, Sammy and Lee, Antony R.},
  title   = {Continuous Variable Quantum Information: {G}aussian States and Beyond},
  journal = {\textnormal{\href{https://doi.org/10.1142/S1230161214400010}{Open Systems \& Information Dynamics 21, 1440001 (2014)}}},
  volume  = {},
  pages   = {},
  year    = {},
  doi     = {10.1142/S1230161214400010}
}

@article{Ferraro2005,
  author  = {Ferraro, Alessandro and Olivares, Stefano and  Paris, Matteo G. A.},
  title   = {Gaussian states in continuous variable
quantum information},
  journal = {\textnormal{\href{https://doi.org/10.48550/arXiv.quant-ph/0503237}{arXiv:quant-ph/0503237 (2005)}}},
  volume  = {},
  pages   = {},
  year    = {},
  doi     = {10.48550/arXiv.quant-ph/0503237}
}

@phdthesis{Hastrup2021,
  author = {Hastrup, Jacob},
  title  = {Continuous-Variable Quantum Codes for Fault-Tolerant Quantum Information Processing},
  school = {\href{https://backend.orbit.dtu.dk/ws/portalfiles/portal/271733143/PhD_Thesis_002_.pdf}
            {Technical University of Denmark, 2021}},
  year   = {}
}

@article{Holevo2007,
  author  = { Holevo, Alexander},
  title   = {One-mode quantum {G}aussian channels: Structure and quantum capacity},
  journal = {\textnormal{\href{https://doi.org/10.1134/S0032946007010012}{Problems of Information Transmission 43, 1 (2007)}}},
  volume  = {},
  pages   = {},
  year    = {},
  doi     = {10.1134/S0032946007010012}
}

@book{Gardiner2004,
  author    = {Gardiner, Crispin W. and Zoller, Peter},
  title     = {Quantum Noise},
  publisher = {\href{https://link.springer.com/book/9783540223016}
               {Springer, 2004}},
  year      = {}
}

@article{Liu2024,
  author  = {Liu, Yu-Hong  and others},
  title   = {Optimal control of linear {G}aussian quantum systems via quantum learning control},
  journal = {\textnormal{\href{https://doi.org/10.1103/PhysRevA.109.063508}{Physical Review A 109, 063508 (2024)}}},
  volume  = {},
  pages   = {},
  year    = {},
  doi     = {10.1103/PhysRevA.109.063508}
}

@article{Mirkhalaf2024,
  author  = { Mirkhalaf, Safoura and Mehboudi, Mohammad and   Nafari Qaleh, Zohre and Rahimi-Keshari, Saleh},
  title   = {Operational significance of nonclassicality in nonequilibrium {G}aussian quantum thermometry},
  journal = {\textnormal{\href{https://doi.org/10.1088/1367-2630/ad23a1}{New Journal of Physics 26, 023046 (2024)}}},
  volume  = {},
  pages   = {},
  year    = {},
  doi     = {10.1088/1367-2630/ad23a1}
}

@article{Ernesto2026,
  author  = {Ernesto Rodríguez Quintela, Frank },
  title   = {Exceptional points in {G}aussian channels: diffusion gauging and drift-governed spectrum},
  journal = {\textnormal{\href{https://doi.org/10.48550/arXiv.2601.16121}{arXiv:2601.16121 (2026)}}},
  volume  = {},
  pages   = {},
  year    = {},
  doi     = {10.48550/arXiv.2601.16121}
}

@article{Massignan2015,
  author  = {Massignan, Pietro and Lampo, Aniello and Wehr, Jan and Lewenstein, Maciej},
  title   = {Quantum {B}rownian motion with inhomogeneous damping and diffusion},
  journal = {\textnormal{\href{https://doi.org/10.1103/PhysRevA.91.033627}{Physical Review A 91, 033627 (2015)}}},
  volume  = {},
  pages   = {},
  year    = {},
  doi     = {10.1103/PhysRevA.91.033627}
}

@article{Ma2018,
  author  = {Ma, Shan and  Woolley, Matthew J. and Petersen, Ian R.},
  title   = {A derivation of moment evolution equations for linear open quantum systems},
  journal = {\textnormal{\href{https://doi.org/10.48550/arXiv.1801.04149}{arXiv:1801.04149 (2018)}}},
  volume  = {},
  pages   = {},
  year    = {},
  doi     = {10.48550/arXiv.1801.04149}
}

@book{Breuer2007,
  author    = {Breuer, Heinz-Peter and  Petruccione, Francesco},
  title     = {The Theory of Open Quantum Systems},
  publisher = {\textnormal{\href{https://doi.org/10.1093/acprof:oso/9780199213900.001.0001}{Oxford University Press, 2007}}},
  address   = {},
  year      = {},
  doi     = {10.1093/acprof:oso/9780199213900.001.0001}
}

@book{Schlosshauer2007,
  author    = {Schlosshauer, Maximilian},
  title     = {Decoherence
and the Quantum-To-Classical Transition},
  publisher = {\textnormal{\href{https://doi.org/10.1007/978-3-540-35775-9}{Springer, 2007}}},
  address   = {},
  year      = {},
  doi     = {10.1007/978-3-540-35775-9}
}

@article{Vasile2011,
  author  = {Vasile, Ruggero and Maniscalco, Sabrina and Paris, Matteo G. A. and Breuer, Heinz-Peter and Piilo, Jyrki},
  title   = {Quantifying non-{M}arkovianity of continuous-variable {G}aussian dynamical maps},
  journal = {\textnormal{\href{https://doi.org/10.1103/PhysRevA.84.052118}{Physical Review A 84, 052118 (2011)}}},
  volume  = {},
  number  = {},
  pages   = {},
  year    = {},
  doi     = {10.1103/PhysRevA.84.052118}
}

@article{Vasile2010,
  author  = {Vasile, Ruggero and Giorda, Paolo and Olivares, Stefano and Paris, Matteo G. A. and Maniscalco, Sabrina},
  title   = {Nonclassical correlations in non-{M}arkovian continuous-variable systems},
  journal = {\textnormal{\href{https://doi.org/10.1103/PhysRevA.82.012313}{Physical Review A 82, 012313 (2010)}}},
  volume  = {},
  number  = {},
  pages   = {},
  year    = {},
  doi     = {10.1103/PhysRevA.82.012313}
}

@article{Chruscinski2011,
  author  = {Chruściński, Dariusz and Kossakowski, Andrzej and  Rivas, Ángel},
  title   = {Measures of non-{M}arkovianity: Divisibility versus backflow of information},
  journal = {\textnormal{\href{https://doi.org/10.1103/PhysRevA.83.052128}{Physical Review A 83, 052128 (2011)}}},
  volume  = {},
  number  = {},
  pages   = {},
  year    = {},
  doi     = {10.1103/PhysRevA.83.052128}
}

@article{Chruscinski2012,
  author  = {Chruściński, Dariusz and Kossakowski, Andrzej},
  title   = {Markovianity criteria for quantum
evolution},
  journal = {\textnormal{\href{https://doi.org/10.1088/0953-4075/45/15/154002}{Journal of Physics B: Atomic, Molecular and Optical Physics 45, 154002 (2012)}}},
  volume  = {},
  number  = {},
  pages   = {},
  year    = {},
  doi     = {10.1088/0953-4075/45/15/154002}
}

@article{Chruscinski2013,
  author  = {Chruściński, Dariusz},
  title   = {Characterizing non-{M}arkovian quantum
evolution},
  journal = {\textnormal{\href{https://doi.org/10.1088/0031-8949/2013/T153/014009}{Physica Scripta T153, 014009 (2013)}}},
  volume  = {},
  number  = {},
  pages   = {},
  year    = {},
  doi     = {10.1088/0031-8949/2013/T153/014009}
}

@article{DAriano2024,
  author  = {D’Ariano,Giacomo and  Paris, Matteo G. A. and  Sacchi, Massimiliano F.},
  title   = {Quantum Tomography},
  journal = {\textnormal{\href{https://doi.org/10.48550/arXiv.quant-ph/0302028}{arXiv:quant-ph/0302028 (2003)}}},
  volume  = {},
  number  = {},
  pages   = {},
  year    = {},
  doi     = {10.48550/arXiv.quant-ph/0302028}
}

@article{DAriano1995,
  author  = {D’Ariano,Giacomo and  Ulf Leonhardt and H. Paul},
  title   = {Homodyne detection of the density matrix of the radiation field
},
  journal = {\textnormal{\href{https://doi.org/10.1103/PhysRevA.52.R1801}{Physical Review A 52, R1801 (1995)}}},
  volume  = {},
  number  = {},
  pages   = {},
  year    = {},
  doi     = {10.1103/PhysRevA.52.R1801}
}

@article{DAriano1999,
  author  = {D’Ariano, Giacomo and  Sacchi, Massimiliano F. and  Kumar, Prem},
  title   = {Universal homodyne tomography with a single local oscillator
},
  journal = {\textnormal{\href{https://doi.org/10.1103/PhysRevA.61.013806}{Physical Review A 61, 013806 (1999)}}},
  volume  = {},
  number  = {},
  pages   = {},
  year    = {},
  doi     = {10.1103/PhysRevA.61.013806}
}

@article{DAriano1994,
  author  = {D’Ariano, Giacomo and  Macchiavello, C. and  Paris, Matteo G. A.},
  title   = {Detection of the density matrix through optical homodyne tomography
without filtered back projection
},
  journal = {\textnormal{\href{https://doi.org/10.1103/PhysRevA.50.4298}{Physical Review A 50, 4298 (1994)}}},
  volume  = {},
  number  = {},
  pages   = {},
  year    = {},
  doi     = {10.1103/PhysRevA.50.4298}
}

@article{Merkel2013,
  author  = {Merkel, Seth T. and others},
  title   = {Self-consistent quantum process tomography
},
  journal = {\textnormal{\href{https://doi.org/10.1103/PhysRevA.87.062119}{Physical Review A 87, 062119 (2013)}}},
  volume  = {},
  number  = {},
  pages   = {},
  year    = {},
  doi     = {10.1103/PhysRevA.87.062119}
}

@article{James2001,
  author  = {James, Daniel F. V. and Kwiat, Paul G. and  Munro, William J. and  White, Andrew G.},
  title   = {Measurement of qubits
},
  journal = {\textnormal{\href{https://doi.org/10.1103/PhysRevA.64.052312}{Physical Review A 64, 052312 (2001)}}},
  volume  = {},
  number  = {},
  pages   = {},
  year    = {},
  doi     = {10.1103/PhysRevA.64.052312}
}

@article{Banaszek1999,
  author  = {Banaszek, Konrad and  D’Ariano, Giacomo and Paris, Matteo G. A. and Sacchi, Massimiliano F.},
  title   = {Maximum-likelihood estimation of the density matrix
},
  journal = {\textnormal{\href{https://doi.org/10.1103/PhysRevA.61.010304}{Physical Review A 61,  010304(R) (1999)}}},
  volume  = {},
  number  = {},
  pages   = {},
  year    = {},
  doi     = {10.1103/PhysRevA.61.010304}
}

@article{Fiurasek2001,
  author  = { Fiurášek, Jaromír and  Hradil, Zdeněk},
  title   = {Maximum-likelihood estimation of quantum processes
},
  journal = {\textnormal{\href{https://doi.org/10.1103/PhysRevA.63.020101}{Physical Review A 63,  020101(R) (2001)}}},
  volume  = {},
  number  = {},
  pages   = {},
  year    = {},
  doi     = {10.1103/PhysRevA.63.020101}
}

@article{Sacchi2001,
  author  = {Sacchi, Massimiliano F.},
  title   = {Maximum-likelihood reconstruction of completely positive maps

},
  journal = {\textnormal{\href{https://doi.org/10.1103/PhysRevA.63.054104}{Physical Review A 63, 054104 (2001)}}},
  volume  = {},
  number  = {},
  pages   = {},
  year    = {},
  doi     = {10.1103/PhysRevA.63.054104}
}

@article{Varona2025,
  author  = { Varona, Santiago and Müller, Markus and  Bermudez, Alejandro },
  title   = {Lindblad-like quantum tomography for
non-{M}arkovian quantum dynamical maps

},
  journal = {\textnormal{\href{https://doi.org/10.1038/s41534-025-01044-7}{npj quantum information 11, 96 (2025)}}},
  volume  = {},
  number  = {},
  pages   = {},
  year    = {},
  doi     = {10.1038/s41534-025-01044-7}
}

@article{White2022,
  author  = {White, Gregory A. L. and  Pollock, Felix A. and  Hollenberg, Lloyd C. L. and  Modi, Kavan and  Hill, Charles D.},
  title   = {Non-{M}arkovian quantum process tomography
},
  journal = {\textnormal{\href{https://doi.org/10.1103/PRXQuantum.3.020344}{PRX Quantum 3, 020344 (2022)}}},
  volume  = {},
  number  = {},
  pages   = {},
  year    = {},
  doi     = {10.1103/PRXQuantum.3.020344}
}

@article{Cao2026,
  author  = {Cao, Hanyan and Feng, Dongyang and Lu, Zixuan and Ye, Cheng  and Pan, Feng},
  title   = {Differentiable maximum likelihood noise estimation for quantum error correction
},
  journal = {\textnormal{\href{https://doi.org/10.48550/arXiv.2602.19722}{arXiv:2602.19722 (2026)}}},
  volume  = {},
  number  = {},
  pages   = {},
  year    = {},
  doi     = {10.48550/arXiv.2602.19722}
}

@article{Smolin2012,
  author  = {Smolin, John A. and Gambetta, Jay M. and  Smith, Graeme},
  title   = {Efficient method for computing the maximum-likelihood quantum state from measurements
with additive {G}aussian noise
},
  journal = {\textnormal{\href{https://doi.org/10.1103/PhysRevLett.108.070502}{Physical Review Letters 108, 070502 (2012)}}},
  volume  = {},
  number  = {},
  pages   = {},
  year    = {},
  doi     = {10.1103/PhysRevLett.108.070502}
}

@article{Tanaka2021,
  author  = {Tanaka, Tomoki and others},
  title   = {Amplitude estimation via maximum likelihood on noisy quantum computer
},
  journal = {\textnormal{\href{https://doi.org/10.1007/s11128-021-03215-9}{Quantum Information Processing 20, 293 (2021)}}},
  volume  = {},
  number  = {},
  pages   = {},
  year    = {},
  doi     = {10.1007/s11128-021-03215-9}
}

@article{Vasile2009,
  author  = {Vasile, Ruggero and Olivares, Stefano and Paris, Matteo G. A. and Maniscalco, Sabrina},
  title   = {Continuous-variable-entanglement dynamics in structured reservoirs},
  journal = {\textnormal{\href{https://doi.org/10.1103/PhysRevA.80.062324}{Physical Review A 80, 062324 (2009)}}},
  volume  = {},
  pages   = {},
  year    = {},
  doi     = {10.1103/PhysRevA.80.062324}
}

@article{Torre2015,
  author  = {Torre, Gianpaolo and Roga, Wojciech and Illuminati, Fabrizio},
  title   = {Non-{M}arkovianity of {G}aussian channels},
  journal = {\textnormal{\href{https://doi.org/10.1103/PhysRevLett.115.070401}{Physical Review Letters 115, 070401 (2015)}}},
  volume  = {},
  pages   = {},
  year    = {},
  doi     = {10.1103/PhysRevLett.115.070401}
}

@article{Souza2015,
  author  = {de Souza, Leonardo A. M. and Dhar, Himadri Shekhar and Bera, Manabendra Nath and Liuzzo-Scorpo, Pietro and Adesso, Gerardo},
  title   = {Gaussian interferometric power as a measure of continuous-variable non-{M}arkovianity},
  journal = {\textnormal{\href{https://doi.org/10.1103/PhysRevA.92.052122}{Physical Review A 92, 052122 (2015)}}},
  volume  = {},
  pages   = {},
  year    = {},
  doi     = {10.1103/PhysRevA.92.052122}
}

@article{Trapani2016,
  author  = {Trapani, Jacopo and Paris, Matteo G. A.},
  title   = {Nondivisibility versus backflow of information in understanding revivals of quantum correlations
for continuous-variable systems interacting with fluctuating environments},
  journal = {\textnormal{\href{https://doi.org/10.1103/PhysRevA.93.042119}{Physical Review A 93, 042119 (2016)}}},
  volume  = {},
  pages   = {},
  year    = {},
  doi     = {10.1103/PhysRevA.93.042119}
}

@article{LiuzzoScorpo2017,
  author  = {Liuzzo-Scorpo, Pietro and Roga, Wojciech and Souza, Leonardo A. M. and Bernardes, Nadja K. and Adesso, Gerardo},
  title   = {Non-{M}arkovianity hierarchy of {G}aussian processes and quantum amplification},
  journal = {\textnormal{\href{https://doi.org/10.1103/PhysRevLett.118.050401}{Physical Review Letters 118, 050401 (2017)}}},
  volume  = {},
  pages   = {},
  year    = {},
  doi     = {10.1103/PhysRevLett.118.050401}
}

@article{Simon1994,
  author  = {Simon, Rajiah and Mukunda, N and Dutta, Biswadeb},
  title   = {Quantum-noise matrix for multimode systems: U(n) invariance, squeezing, and normal forms},
  journal = {\textnormal{\href{https://doi.org/10.1103/PhysRevA.49.1567}{Physical Review A 49, 1567 (1994)}}},
  volume  = {},
  pages   = {},
  year    = {},
  doi     = {10.1103/PhysRevA.49.1567}
}

@article{Naeij2020,
  author  = {Soltanmanesh, Ali and Naeij, Hamid Reza and Shafiee, Afshin},
  title   = {Can thermodynamic behavior of {A}lice’s particle affect {B}ob’s particle?},
  journal = {\textnormal{\href{https://doi.org/10.1038/s41598-020-65800-9}{Scientific Reports 10, 1 (2020)}}},
  volume  = {},
  pages   = {},
  year    = {},
  doi     = {10.1038/s41598-020-65800-9}
}

@article{Naeij2025,
  author  = {Naeij, Hamid Reza},
  title   = {Open quantum system approaches to superconducting qubits},
  journal = {\textnormal{\href{https://doi.org/10.1007/s11128-025-04838-y}{Quantum Information Processing 24, 1 (2025)}}},
  volume  = {},
  pages   = {},
  year    = {},
  doi     = {10.1007/s11128-025-04838-y}
}

@article{Toscano2021,
  author  = {Toscano, Fabricio and Bosyk, Gustavo M. and Zozor, Steeve and Portesi, Mariela},
  title   = {Fluctuation and dissipation in memoryless open quantum evolutions},
  journal = {\textnormal{\href{https://doi.org/10.1103/PhysRevA.104.062207}{Physical Review A 104, 062207 (2021)}}},
  volume  = {},
  pages   = {},
  year    = {},
  doi     = {10.1103/PhysRevA.104.062207}
}

@article{Toscano2022,
  author  = {Toscano, Fabricio and Nicacio, Fernando},
  title   = {Thermal equilibrium in {G}aussian dynamical semigroups},
  journal = {\textnormal{\href{https://doi.org/10.1103/PhysRevA.106.062207}{Physical Review A 106, 062207 (2022)}}},
  volume  = {},
  pages   = {},
  year    = {},
  doi     = {10.1103/PhysRevA.106.062207}
}

@article{Gardiner1985,
  author  = {Gardiner, Crispin  and Collett, Matthew },
  title   = {Input and output in damped quantum systems: Quantum stochastic differential equations and the master equation},
  journal = {\textnormal{\href{https://doi.org/10.1103/PhysRevA.31.3761}{Physical Review A 31, 3761 (1985)}}},
  volume  = {},
  pages   = {},
  year    = {},
  doi     = {10.1103/PhysRevA.31.3761}
}

@article{Garraway1997,
  author  = {Garraway, Barry M.},
  title   = {Nonperturbative decay of an atomic system in a cavity},
  journal = {\textnormal{\href{https://doi.org/10.1103/PhysRevA.55.2290}{Physical Review A 55, 2290 (1997)}}},
  volume  = {},
  pages   = {},
  year    = {},
  doi     = {10.1103/PhysRevA.55.2290}
}

@article{Dalton2001,
  author  = {Dalton, Bryan J. and Barnett, Stephen M. and Garraway, Barry M.},
  title   = {Theory of pseudomodes in quantum optical processes},
  journal = {\textnormal{\href{https://doi.org/10.1103/PhysRevA.64.053813}{Physical Review A 64, 053813 (2001)}}},
  volume  = {},
  pages   = {},
  year    = {},
  doi     = {10.1103/PhysRevA.64.053813}
}

@article{Pleasance2020,
  author  = {Pleasance, Graeme and Garraway, Barry M. and Petruccione, Francesco},
  title   = {Generalized theory of pseudomodes for exact descriptions of non-{M}arkovian quantum processes},
  journal = {\textnormal{\href{https://doi.org/10.1103/PhysRevResearch.2.043058}{Physical Review Research 2, 043058 (2020)}}},
  volume  = {},
  pages   = {},
  year    = {},
  doi     = {10.1103/PhysRevResearch.2.043058}
}

@article{Pleasance2021,
  author  = {Pleasance, Graeme and Petruccione, Francesco},
  title   = {Pseudomode description of general open quantum system dynamics: non-perturbative master equation for the spin-boson model},
  journal = {\textnormal{\href{https://doi.org/10.48550/arXiv.2108.05755}{arXiv:2108.05755 (2021)}}},
  volume  = {},
  pages   = {},
  year    = {},
  doi     = {10.48550/arXiv.2108.05755}
}

@article{Clason2021,
  author  = {Christian Clason},
  title   = {Regularization of Inverse Problems},
  journal = {\textnormal{\href{https://doi.org/10.48550/arXiv.2001.00617}{arXiv:2001.00617 (2021)}}},
  volume  = {},
  pages   = {},
  year    = {},
  doi     = {10.48550/arXiv.2001.00617}
}

@article{Torre2018,
  author  = {Torre, Gianpaolo and Illuminati, Fabrizio},
  title   = {Exact non-{M}arkovian dynamics of {G}aussian quantum channels: {F}inite-time and asymptotic regimes},
  journal = {\textnormal{\href{https://doi.org/10.1103/PhysRevA.98.012124}{Physical Review A 98, 012124 (2018)}}},
  volume  = {},
  pages   = {},
  year    = {},
  doi     = {10.1103/PhysRevA.98.012124}
}

@article{Akaike1974,
  author  = {Akaike, Hirotugu},
  title   = {A new Look at the statistical model identification},
  journal = {\textnormal{\href{https://doi.org/10.1109/TAC.1974.1100705}{IEEE transactions on automatic control 19, 716 (1974)}}},
  volume  = {},
  pages   = {},
  year    = {},
  doi     = {10.1109/TAC.1974.1100705}
}

@article{Schwarz1978,
  author  = {Schwarz,Gideon},
  title   = {Estimating the dimension of a model},
  journal = {\textnormal{\href{https://doi.org/10.1214/aos/1176344136}{The annals of statistics 6, 461 (1978)}}},
  volume  = {},
  pages   = {},
  year    = {},
  doi     = {10.1214/aos/1176344136}
}

@article{Dziak2020,
  author  = {Dziak,John J. and Coffman, Donna L. and Lanza, Stephanie T. and Li,Runze  and Jermiin,Lars S.},
  title   = {Sensitivity and specificity of information criteria},
  journal = {\textnormal{\href{https://doi.org/10.1093/bib/bbz016}{Briefings in bioinformatics 21, 553 (2020)}}},
  volume  = {},
  pages   = {},
  year    = {},
  doi     = {10.1093/bib/bbz016}
}

@article{Guta2012,
  author  = {Gutǎ, Mǎdǎlin and Kypraios, Theodore and Dryden, Ian},
  title   = {Rank-based model selection for multiple ions quantum tomography},
  journal = {\textnormal{\href{https://doi.org/10.1088/1367-2630/14/10/105002}{New Journal of Physics 14, 105002 (2012)}}},
  volume  = {},
  pages   = {},
  year    = {},
  doi     = {10.1088/1367-2630/14/10/105002}
}

@article{Carlesso2021,
  title = {Perturbative algorithm for rotational decoherence},
  author = {Carlesso, Matteo and Naeij, Hamid Reza and Bassi, Angelo},
  journal = {\textnormal{\href{https://doi.org/10.1103/PhysRevA.103.032220}{Physical Review A 103, 032220 (2021)}}},
  volume = {},
  number = {},
  pages = {},
  year = {},
  publisher = {American Physical Society},
  doi = {10.1103/PhysRevA.103.032220}
}

@article{Naeij2022RotationalPhoton,
  title        = {Rotational decoherence due to thermal photon scattering},
  author       = {Naeij, Hamid Reza},
  journal      = {\textnormal{\href{https://doi.org/10.48550/arXiv.2210.06133}{ arXiv:2210.06133 (2022)}}},
  year         = {},
  eprint       = {2210.06133},
  archivePrefix = {arXiv},
  primaryClass = {quant-ph},
  doi     = {10.48550/arXiv.2210.06133},
  volume = {},
  pages = {},
}

@article{Tirandaz2020Cat,
  title     = {Bringing Schr{\"o}dinger's cat to life with non-equilibrium respiration},
  author    = {Tirandaz, Arash and Naeij, Hamid Reza and Shafiee, Afshin},
  journal   = {\textnormal{\href{https://doi.org/10.1088/2040-8986/ab8070}{Journal of Optics 22, 055702 (2020)}}},
  volume    = {},
  pages     = {},
  year      = {},
  doi       = {10.1088/2040-8986/ab8070}
}

@article{mogilevtsev2013cross,
  title     = {Cross-Validated Tomography},
  author    = {Mogilevtsev, Dmitri and Hradil, Zdenek and Rehacek, Jaroslav and Shchesnovich, Valery S.},
  journal   = {\textnormal{\href{https://doi.org/10.1103/PhysRevLett.111.120403}{Physical Review Letters 111, 120403 (2013)}}},
  volume    = {},
  pages     = {},
  year      = {},
  doi       = {10.1103/PhysRevLett.111.120403}
}

@article{tiunov2020experimental,
  title     = {Experimental Quantum Homodyne Tomography via Machine Learning},
  author    = {Tiunov, Egor S. and Tiunova, Vera V. and Ulanov, Alexander E. and Lvovsky, Alexander and Fedorov, Aleksey },
  journal   = {\textnormal{\href{https://doi.org/10.1364/OPTICA.389482}{Optica 7, 448 (2020)}}},
  volume    = {},
  number    = {},
  pages     = {},
  year      = {},
  doi       = {10.1364/OPTICA.389482}
}

@article{Rudolph2025Energy,
  title   = {Energy, Bosons and Computational Complexity},
  author  = {Rudolph, Dorian and others},
  journal = {\textnormal{\href{https://doi.org/10.48550/arXiv.2510.08545}{arXiv:2510.08545 (2025)}}},
  year    = {},
  doi     = {10.48550/arXiv.2510.08545},
  volume = {},
  pages = {},
}

@article{Kossakowski1,
  author  = {Kossakowski, Andrzej},
  title   = {On quantum statistical mechanics of
             non-{Hamiltonian} systems},
  journal = {\textnormal{\href{https://doi.org/10.1016/0034-4877(72)90010-9}{Reports on Mathematical Physics 3, 247 (1972)}}},
  volume  = {},
  number  = {},
  pages   = {},
  year    = {},
  doi     = {10.1016/0034-4877(72)90010-9}
}

@article{Kossakowski2,
  author  = {Kossakowski, Andrzej and Rebolledo, Rolando},
  title   = {On the structure of generators for {non-Markovian} master equations},
  journal = {\textnormal{\href{https://doi.org/10.1142/S1230161209000190}{Open Systems \& Information Dynamics 16, 259 (2009)}}},
  volume  = {},
  number  = {},
  pages   = {},
  year    = {},
  doi     = {10.1142/S1230161209000190}
}

@book{Walls2008,
  author    = {Walls, Daniel Frank  and Milburn, Gerard J.},
  title     = {Quantum Optics},
  publisher = {\textnormal{\href{https://doi.org/10.1007/978-3-540-28574-8}{Springer, 2008}}},
  year      = {},
  doi     = {10.1007/978-3-540-28574-8}
}

@article{Milz2019,
  author  = {Milz, Simon and Kim, Myungshik and
             Pollock, Felix A. and Modi, Kavan},
  title   = {Completely Positive Divisibility Does Not Mean
             {Markovianity}},
  journal = {\textnormal{\href{https://doi.org/10.1103/PhysRevLett.123.040401}{Physical Review Letters 123, 040401 (2019)}}},
  volume  = {},
  number  = {},
  pages   = {},
  year    = {},
  doi     = {10.1103/PhysRevLett.123.040401}
}

@article{Hall2014,
  author  = {Hall, Michael J. W. and Cresser, James D. and
             Li, Li and Andersson, Erika},
  title   = {Canonical form of master equations and
             characterization of non-{Markovianity}},
  journal = {\textnormal{\href{https://doi.org/10.1103/PhysRevA.89.042120}{Physical Review A 89, 042120 (2014)}}},
  volume  = {},
  number  = {},
  pages   = {},
  year    = {},
  doi     = {10.1103/PhysRevA.89.042120}
}

@article{Renault2023,
  author  = {Renault, Paul and others},
  title   = {Experimental Optical Simulator of Reconfigurable
             and Complex Quantum Environment},
  journal = {\href{https://doi.org/10.1103/PRXQuantum.4.040310}
             {\textnormal{PRX Quantum \textbf{4}, 040310 (2023)}}}
}

@article{Aguiar2025,
  author = {Aguiar, Diogo and Wold, Kristian
            and Denisov, Sergey and Ribeiro, Pedro},
  title = {Quantum Liouvillian Tomography},
  journal = {\href{https://arxiv.org/abs/2504.10393}
                  {\textnormal{arXiv:2504.10393 (2025)}}}
}

@article{Reddy2025,
  author = {Reddy, Sohail},
  title = {Characterizing Non-Markovian Dynamics
           of Open Quantum Systems},
  journal = {\href{https://arxiv.org/abs/2503.22147}
             {\textnormal{arXiv:2503.22147 (2025)}}}
}

@article{Guan2026,
  author = {Guan, Vincent and others},
  title = {Identifying Drift, Diffusion, and Causal
           Structure from Temporal Snapshots},
  journal = {\href{https://www.jmlr.org/papers/v27/25-0541.html}
             {\textnormal{J. Mach. Learn. Res.
             \textbf{27}, 1(2026)}}}
}

\end{document}